\documentclass[10pt]{iopart}

\usepackage{pslatex}
\usepackage{iopams}
\usepackage{subfigure}
\usepackage[dvips]{graphicx}
\usepackage{cite}
\usepackage{pdflscape}

\begin{document}
\title[TwoSpect all-sky gravitational wave search]{An all-sky search algorithm for continuous gravitational waves from spinning neutron stars in binary systems}
\author{E~Goetz$^{1,2}$ and K~Riles$^{1}$}
\address{$^1$ University of Michigan, Ann Arbor, MI 48109, USA} 
\address{$^2$ Max-Planck-Institut f\"ur Gravitationphysik, Callinstr. 38, 30167 Hannover, Germany}

\eads{\mailto{evan.goetz@aei.mpg.de} and \mailto{kriles@umich.edu}}

\begin{abstract}
Rapidly spinning neutron stars with non-axisymmetric mass distributions are expected to generate quasi-monochromatic continuous gravitational waves. While many searches for unknown, isolated spinning neutron stars have been carried out, there have been no previous searches for unknown sources in binary systems. Since current search methods for unknown, isolated neutron stars are already computationally limited, expanding the parameter space searched to include binary systems is a formidable challenge. We present a new hierarchical binary search method called TwoSpect, which exploits the periodic orbital modulations of the continuous waves by searching for patterns in doubly Fourier-transformed data. We will describe the TwoSpect search pipeline, including its mitigation of detector noise variations and corrections for Doppler frequency modulation caused by changing detector velocity. Tests on Gaussian noise and on a set of simulated signals will be presented.
\end{abstract}


\section{Introduction}\label{sec:twospectintro}
Searches for neutron stars emitting quasi-monochromatic, continuous gravitational waves using the LIGO, Virgo, GEO600 or TAMA interferometers' data have been carried out for over a decade~\cite{S1knownPulsarSearch, S2knownPulsarSearch, S2Hough, S2cwSearch, S4knownPulsarSearch, S4PSH, S4EatH, EarlyS5PowerFlux, EarlyS5EatH, CasA, S5knownPulsarSearch, TamaCW, VelaVirgo}. Other searches have been performed using prototype interferometers~\cite{LivasThesis, ZuckerThesis, Niebauer1993} and by so-called bar detectors~\cite{HoughNature, AllegroAllsky, AstoneExplorer}. The different search strategies for detecting this type of gravitational radiation can be classified into three general categories: a targeted search for a known neutron star, a directed search at a particular sky location, or a search for unknown neutron stars over the entire sky.

For the first type of search, a known neutron star (typically observed as a radio-pulsar), either isolated or in a binary system, with a well-known ephemeris describing the rotation of the source, can be targeted using methods searching over a very narrow range of parameters~\cite{S1knownPulsarSearch, S2knownPulsarSearch, S4knownPulsarSearch, S5knownPulsarSearch, VelaVirgo, HoughNature}. The second strategy is one in which a search for continuous gravitational waves from a particular sky location that contains a potential source--or many sources--of gravitational wave signals (e.g., Sco X-1, the supernova remnant Cassiopeia A, the galactic center, or globular clusters) are targeted by algorithms that search over a wider range of parameters~\cite{CasA, AstoneExplorer}. Unfortunately, these two strategies are too computationally costly to be used to search over the entire sky for unknown sources, given current computational resources. The third approach attempts to cover a wide region of parameter space, over the entire sky, using computationally efficient analysis algorithms. These methods are intrinsically less sensitive than the targeted search algorithms, but are computationally tractable~\cite{S2Hough, S4PSH, S4EatH, EarlyS5PowerFlux, EarlyS5EatH, AllegroAllsky, ZuckerThesis}.

Even these current all-sky search algorithms, however, are not designed to search for unknown neutron stars in binary systems. The continuous gravitational wave signal from a binary source is frequency-modulated by the source's orbital motion, which requires, in general, five (non-relativistic) unknown parameters to describe~\cite{Vecc2001}. Since present all-sky search algorithms cannot cope with the increased computational cost of searching over these additional parameters, one must find alternative methods to reduce the dimensionality of the parameter space to search and to evaluate trade-offs between search sensitivity and computational efficiency.

We have developed such a method, one that detects the periodic Doppler shift the signal frequency experiences with each orbit of the binary system. Our search method is able to exploit the regular periodicity of these extremely stable orbits over the course of the observation time ($T_{obs}\sim1$ year) using two successive, computationally efficient Fourier transformations of the detector output. Hence, the name of the algorithm: {\it TwoSpect} (from the use of two successive spectral transformations). It should be noted, the present TwoSpect algorithm restricts the orbits to near circularity, where only three parameters are searched over.

More than half of the observed radio pulsars with rotation rates that could plausibly emit gravitational waves in the most sensitive band of the LIGO detectors are located in binary systems. Moreover, most of these binary systems have very small eccentricities. It is thus prudent to develop all-sky binary search techniques, optimized for circular orbits, to complement the mature techniques now used to search for isolated neutron stars.

\section{Astrophysical parameter space}\label{sec:twospectparameterspace}
Although observational evidence has shown that neutron stars exist within binary systems with a wide variety of binary orbital parameters~\cite{atnfcatalog,LMXBcat}\footnote{The Australian National Telescope Facility maintains a database of all known radio pulsars at http://www.atnf.csiro.au/research/pulsar/psrcat/.}, in many cases, especially for systems with neutron star spin frequencies in the LIGO search band, these systems are nearly circularized (eccentricity $e\lesssim10^{-3}$). To gain a sense of the largest scale of Doppler shift that occurs due to the binary orbit, which affects the parameter space to be searched, we calculate the velocity of the neutron star based on its motion about the binary system's center of mass (using the Newtonian approximation)
\begin{equation}
 \frac{GM_{\rm NS}^2q}{(r_1+r_2)^2} = M_{\rm NS}\Omega^2r_1\,,
\end{equation}
where $G$ is the gravitational constant, $M_{\rm NS}$ is the neutron star mass, $q\equiv M_2/M_{\rm NS}$ is the ratio of the companion object to the neutron star mass, $\Omega$ is the angular frequency of the orbiting bodies, and $r_1$ and $r_2$ are the distances from the center of the neutron star and companion star to the binary system center-of-mass, respectively. Then, using the period of the binary orbit, $P=2\pi/\Omega$, and the velocity of the neutron star, $v_{\rm NS}=2\pi r_1/P$, we solve for the neutron star velocity:
\begin{equation}
 v_{\rm NS} = \left(\frac{2\pi G M_{\rm NS}}{P}\right)^{1/3} \left[\frac{q}{(1+q)^{2/3}}\right]\,.
\end{equation}
Therefore, the maximum Doppler shift observable (the binary system is observed edge-on) $\Delta f_{\rm max} = f v_{\rm max}/c$, where $f$ is the source frequency of gravitational waves and $c$ is the speed of light, will be
\begin{equation}
\Delta f_{\rm max} \simeq 1.82 \, \left(\frac{f}{1\,{\rm kHz}}\right) \left(\frac{M_{\rm NS}}{1.4\,M_{\odot}}\right)^{1/3} \left(\frac{P}{2\,{\rm h}}\right)^{-1/3} \left[\frac{q}{(1+q)^{2/3}}\right]\,{\rm Hz}.
\label{eq:dfmaxastro}
\end{equation}
Thus, the gravitational wave signal in the Solar System barycenter (SSB) frame will take the form $h_{\rm SSB}(t) = h_+[\Phi(t)] + h_\times[\Phi(t)]$, where $h_+$ and $h_\times$ are the two gravitational wave polarizations~\cite{S2Hough}. To first order in phase when the source is, for example, edge-on to the SSB
\begin{equation}
\Phi(t) = \Phi_0 + 2\pi\{f_0 + \Delta f_{\rm max}\sin[\Omega (T-T_0) + \phi_0]\}(T-T_0)\,,
\end{equation}
where $T$ is the time in the SSB frame and $\Phi_0$, $f_0$, $T_0$, and $\phi_0$ are the initial phase, frequency, start time in the SSB frame, and orbital phase at the start of the observation, respectively.

The maximum Doppler shift can vary by orders of magnitude depending on the intrinsic spin frequency of the neutron star and the mass of the companion star.  The current implementation of the TwoSpect algorithm is primarily aimed at detecting neutron stars in low eccentricity orbits ($e<0.1$), and is targeting signals with an observable frequency modulation of $\Delta f_{\rm obs}\leq1$~Hz with a lower bound on the search dependent on the coherence time of the initial Fourier transforms.

The observed amplitude of Doppler shift caused by the orbital motion also depends on the inclination angle $i$ of the orbital system to the line of sight between the SSB and the source. From the observer's point of view, the Doppler shift is scaled by $\Delta f_{\rm obs} = \Delta f_{\rm max} \sin i$, where $i$ is the inclination angle of the binary orbital plane with respect to the vector pointing from the detector to the sky position. The observed Doppler shift is thus coupled to the parameters
\begin{equation}
 \Delta f_{\rm obs} \propto M_{\rm NS}^{1/3} \frac{q}{(1+q)^{2/3}} \sin i\,.
\end{equation}
That is, the constituent masses of the binary system, $M_{\rm NS}$ and $q$, and the inclination angle of the system to the observer, $i$, cannot be separately determined from the observed Doppler shift alone.

The range of binary orbital periods to search depends on several factors, and is described in more detail in section~\ref{sec:twospectdetails}. In summary, the number of neutron star orbits during the observation time determines the upper bound of the search over the orbital period. Simulations show that a reasonable upper bound on the orbital period range is one-fifth of the total observation time, $T_{\rm obs}$. Meanwhile, the lower bound is governed by the coherence time of the initial Fourier transformation used to cover the parameter space and by the highest frequency to be searched. From this limit, the shortest period is $P_{\rm min}=2$~h.

The frequency band to be searched is determined by the sensitivity of the detectors to gravitational waves. In this case, the LIGO detectors are most sensitive in the range of $50\leq f \leq1000$~Hz, with the best strain sensitivity occurring near 150~Hz. A neutron star with a non-axisymmetric crust is expected to emit gravitational radiation at a frequency twice its rotational frequency; hence stars with spin frequencies of $25\leq \nu \leq500$~Hz are of most interest. There are presently 182 known pulsars with observed spin frequencies greater than 25~Hz, and, of these pulsars, 111 are located within binary systems~\cite{atnfcatalog}.

\section{Overview of the TwoSpect analysis technique}\label{sec:twospectmethod}
As described in section~\ref{sec:twospectintro}, the TwoSpect algorithm exploits the long-term periodicity of signal power within a range of frequency bins of sequential, short coherence length Fourier transforms (so-called {\it SFTs}). As in other all-sky, semi-coherent LIGO search methods, the typical coherence time for SFTs is $T_{\rm SFT}=1800$~s~\cite{HoughSearch,S2Hough}, but TwoSpect also uses shorter coherence lengths as well (see section~\ref{sec:twospectdetails}). While the previously published all-sky searches have allowed arbitrary gaps between SFTs, the algorithm presented here requires SFT start times to be separated by integer multiples of their common coherence time. Drop-outs in the data stream, either due to loss of detector control or periods of poor data quality, are filled with zeroes to maintain synchronization.

\begin{figure}
    \begin{center}
      \includegraphics[width=0.83\textwidth]{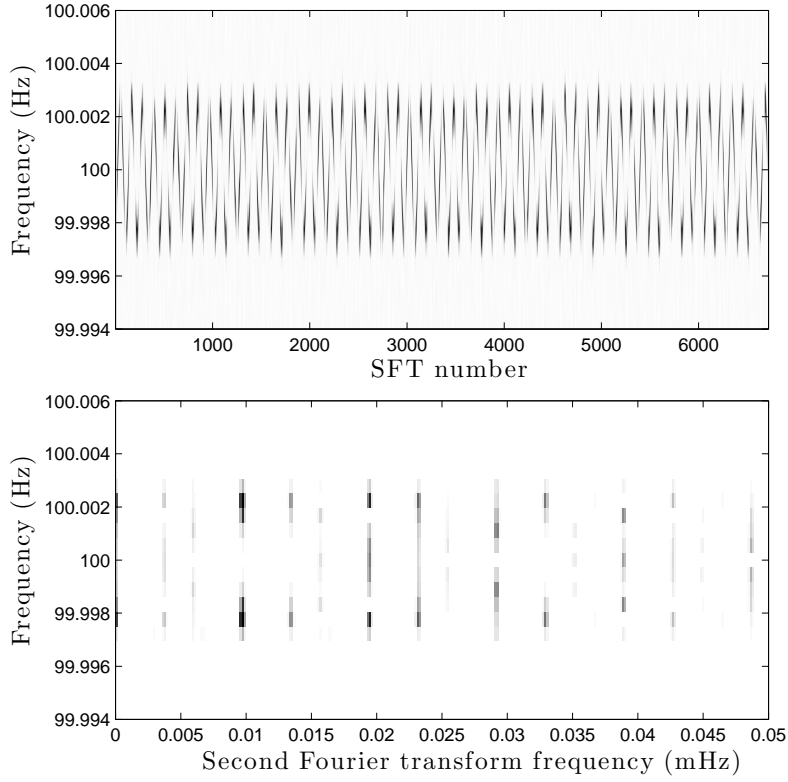}
      \caption{Top: Time-frequency plot of a simulated strong continuous wave signal in detector data over 10 weeks of observation. The SFT data ($T_{\rm SFT}=1800$~s) has been corrected for the motion of the detector and the antenna pattern weighting, assuming a signal with circular polarization. Bottom: After Fourier transformation of each frequency bin's powers as a function of time, the periodicity of the signal is clearly visible with harmonics of the binary orbital period clearly evident in the second Fourier transform. Dark pixels correspond to increased power, and in each plot the darkest pixels correspond to pixels with 50\% or higher power relative to the highest power pixel in the plot.}
      \label{fig:TFandFFplot}
    \end{center}
\end{figure}

The magnitude-squared of the Fourier coefficients (the ``power'') calculated from the SFTs produced from the calibrated detector gravitational wave channel, $h(t)$, are computed in the detector's rest frame and must be shifted to account for the motion of the detector with respect to the SSB (see section~\ref{sec:twospectdetails} for further description). After this correction is made, a second Fourier transform is computed for each SFT frequency bin power as a function of time. Signals with periodically varying frequency will cause excess power to be found in the second Fourier transforms' frequency bins corresponding to the fundamental orbital frequency and, typically, the higher harmonic frequencies. Figure~\ref{fig:TFandFFplot} shows an example of a strong, periodically varying signal from a simulated source of continuous gravitational waves located in a binary system as it would be observed by a LIGO detector. Note the modulation of the signal power in the SFTs (top plot) due to the time-varying antenna pattern of a LIGO detector. Also observe that the fundamental frequency $1/P=9.755$~$\mu$Hz and higher harmonics are clearly evident as the strong pixel powers in the bottom plot. The example shown is characteristic of the type of signal this search algorithm is targeting.

Once the second Fourier transform powers are computed, the algorithm must locate the pixels with excess power and obtain the most likely set of signal parameters. For TwoSpect, these signal parameters are $(\alpha,\delta,f,P,\Delta f)$ where $(\alpha,\delta)$ is the sky location in right ascension and declination, $f$ is the frequency of the gravitational wave signal in the SSB frame, $P$ is the orbital period of the detected signal, and $\Delta f$ is the observed frequency modulation amplitude. One can also include potential spin-down (or spin-up), $\dot{f}$, of the neutron star as well, but for now, we neglect the intrinsic spin-down of the neutron star. The observed spin-down of millisecond pulsars in binary systems is typically much smaller than isolated pulsars ($|\dot{f}|\lesssim10^{-16}$~Hz~s$^{-1}$)~\cite{atnfcatalog}. Implementation of a spin-down parameter search would increase the TwoSpect computational cost. 

By reducing the binary orbital search parameters from five for the general binary orbit (three for a circularized orbit) to the two parameters used by TwoSpect, computational efficiency is gained at the cost of sensitivity. The computational savings provided permit the search to complete in a reasonable amount of time. This approach is robust against phase variations due to accretion which is likely occurring in many binary systems.

\section{Details of the TwoSpect algorithm}\label{sec:twospectdetails}
\subsection{TwoSpect parameter space}\label{sec:paramspace}
\begin{figure}
 \begin{center}
  \includegraphics[width=0.7\textwidth]{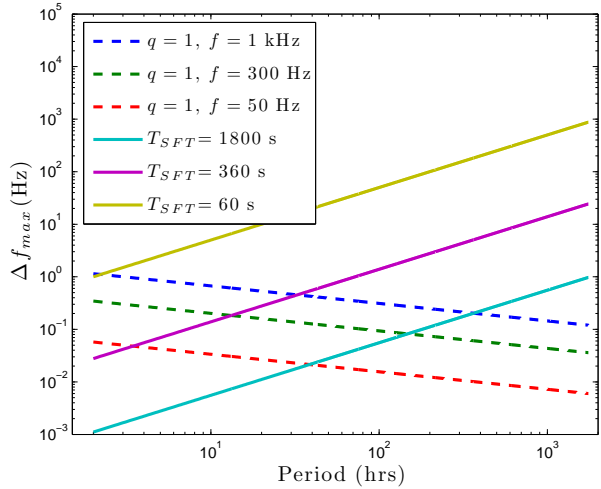}
  \caption{The maximum frequency modulation $\Delta f_{\rm max}$ from equation~\ref{eq:dfmaxastro} for three different signal frequencies, 50 Hz, 300 Hz, and 1 kHz (all with $q=1$), are shown as dashed lines. The maximum frequency modulation for a given period used by the TwoSpect algorithm is shown in solid lines, from equation~\ref{eq:TsftLim}, with successively shorter $T_{\rm SFT}$ values used in order to illustrate the extent of the parameter space covered.}
  \label{fig:dFvPwithTcoh}
 \end{center}
\end{figure}

The calibrated time series strain data, $h(t)$, from a detector is divided into short segments of length $T_{\rm SFT}$ that are coherently analyzed using the FFTW Fourier transform algorithm~\cite{FFTW}. These short stretches of data are windowed using the Hann window function in order to minimize signal leakage into neighboring frequency bins, and each SFT segment overlapped by 50\%. An overlap of 50\% corresponds to the amount that the time-series segment from which the SFT data is produced is shifted for each adjacent SFT produced. This analysis assumes the signal is Doppler modulated by the source motion in such a way that it moves periodically among SFT frequency bins. In order to constrain the signal primarily to a single frequency bin for a single SFT, the SFT coherence time is bounded by
\begin{equation}
 T_{\rm SFT} \leq \left(\frac{P}{2\Delta f_{\rm obs}}\right)^{1/2}\,.
 \label{eq:TsftLim}
\end{equation}
Using equation~\ref{eq:TsftLim}, different regions of parameter space are probed via different coherence times for the SFTs (see figure~\ref{fig:dFvPwithTcoh}).

The best sensitivity is achieved when using the longest coherence time possible for the SFTs. For each choice of SFT coherence time, the parameter space is selected to give the best possible sensitivity. The algorithm allows for different choices of SFT coherence times in order to cover the wide range of binary orbital parameters. The parameter space covered by each choice of SFT coherence time is the region below the solid lines in figure~\ref{fig:dFvPwithTcoh}.

\subsection{Data preparation}\label{sec:dataprep}
Only the calibrated detector $h(t)$ data for which the detector was operating in its nominal data-taking condition is analyzed. The data is divided into segments prior to Fourier analysis, with each segment having length of $T_{\rm SFT}$. Each segment begins at integer multiples of the SFT coherence time. Gaps in the SFT segments (due to poor detector operating conditions or the detector being off-line) are filled with zeroes.

Since the detector is located in an accelerating reference frame with respect to the binary system, the detector data is corrected to mimic observations in the inertial frame of the SSB. The instantaneous signal frequency in the SSB frame, $\hat{f}(t)$, is a frequency-modulated signal in the detector reference frame, $f_{\rm obs}(t)$. The detector signal can be converted to the instantaneous frequency in the SSB frame by~\cite{S2Hough,S4PSH}
\begin{equation}
 f_{\rm obs}(t) - \hat{f}(t) = \hat{f}(t)\frac{\mathbf{v}(t)\cdot\mathbf{\hat{n}}}{c}\,,
\end{equation}
where $\mathbf{v}(t)$ is the detector velocity with respect to the SSB frame and $\mathbf{\hat{n}}$ is the unit vector in the direction of the sky location to be observed. The detector velocity is computed using software barycentering routines based on Earth-Sun ephemeris files\footnote{The Jet Propulsion Laboratory maintains the ephemerides for Solar System bodies; these ephemerides can be found at http://ssd.jpl.nasa.gov/.}~\cite{LALrepository}.

\begin{figure}
 \begin{center}
  \includegraphics[width=0.7\textwidth]{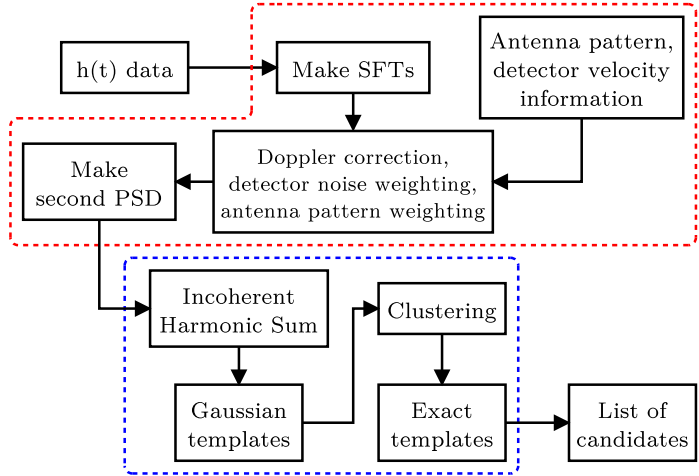}
  \caption{Flow chart schematic illustrating the basic hierarchical TwoSpect search pipeline. The upper portion enclosed is the data-preparation stage of the pipeline, while the lower enclosed portion is the search stage of the pipeline.}
  \label{fig:twospectpipeline}
 \end{center}
\end{figure}

Thus, the powers in frequency bins can be corrected (shifted) for the motion of the detector located on Earth with respect to a given sky location (see figure~\ref{fig:twospectpipeline}). This type of ``barycentering'' is used in many semi-coherent, all-sky search algorithms~\cite{S4PSH,EarlyS5PowerFlux}. The intrinsic angular step size between two different sky location templates is approximately
\begin{equation}
 \varphi_{\rm min} \approx \frac{c}{(v\sin\theta)_{\rm max}}\,\frac{1}{2fT_{\rm SFT}}\,,
 \label{eq:skyphimin}
\end{equation}
where $v$ is the magnitude of the detector velocity, $\theta$ is the angle between the detector velocity and the unit vector which points to the sky position of the source in the SSB frame, and $f$ is the observation frequency. The number of sky locations searched for a fixed frequency is approximately
\begin{equation}
 N_{\rm sky} \approx 2\times10^4 \left(\frac{f}{100\,{\rm Hz}}\right)^2 \left(\frac{T_{\rm SFT}}{1800\,{\rm s}}\right)^2 \,,
\end{equation}
where the sky is oversampled by using $\varphi_{\rm min}$ everywhere, regardless of detector velocity direction.

For each sky position, several steps are carried out prior to computing the second Fourier transform: (1) the powers of the SFT bins are shifted to correct for the detector velocity; (2) the expectation values of the SFT powers in the absence of a signal are subtracted from each measured SFT power; and (3) the data is weighted for detector antenna pattern variation and for the variance of each SFT in the absence of a signal. The expectation value of power is computed from a running median over SFT frequency bins in order to avoid biasing the background estimate due to instrumental lines or gravitational wave signals. The running median is converted to a mean using the correct bias factor for an exponential distribution~\cite{S4PSH, MedianEstimate, MedianEstimate2}.

The noise-weighted, mean subtracted power in frequency bin $k$ as a function of SFT number $n=1\ldots N$ is taken to be
\begin{equation}
 P_k^{\prime\,n} = \frac{P_k^n - \langle P_k\rangle^n}{(\langle P_k\rangle^n)^2}\left[\sum_{n^\prime=1}^N\frac{1}{(\langle P_k\rangle^{n^\prime})^2}\right]^{-1}\,,
 \label{eq:Pnoiseweightingmeansubtraction}
\end{equation}
where $\langle P_k\rangle^n$ is the expected noise-only power in frequency bin $k$ for SFT $n$, and the term in square brackets is used to normalize the weighting. The original SFT powers, $P_k^n$, are normalized such that the expectation value of random, Gaussian, white noise is equal to 1. The signal-to-noise ratio of $P_k^{\prime\,n}$ can be improved by using our knowledge of the interferometer antenna pattern variations. Including antenna pattern weighting in equation~\ref{eq:Pnoiseweightingmeansubtraction} yields
\begin{equation}
\widetilde{P}_k^n = \frac{F_n^2(P_k^n - \langle P_k\rangle^n)}{(\langle P_k\rangle^n)^2}\left[\sum_{n^\prime}^N\frac{F_{n^\prime}^4}{(\langle P_k\rangle^{n^\prime})^2}\right]^{-1}\,,
\label{eq:Pmaster}
\end{equation}
where $\widetilde{P}_k^n$ are the new values of powers after mean subtraction and noise and antenna pattern weighting, and the antenna pattern is $F_n^2 = F_{n,+}^2 + F_{n,\times}^2$ for a given sky location, interferometer and mid-point of the SFT coherence time. This quantity is equivalent to the variable the PowerFlux search~\cite{S4PSH,EarlyS5PowerFlux} calculates for a circularly polarized gravitational wave, except here we subtract the expected noise so that the expectation value of $\widetilde{P}_k^n$ is equal to zero. The advantage in using this weighting scheme is that SFTs with high noise levels or low antenna pattern values are suppressed, while SFTs with lower noise levels or higher antenna pattern values are more heavily weighted.

The Fourier transform of equation~\ref{eq:Pmaster} is then computed for each frequency bin $k$, and normalized such that the expectation value of the noise is equal to 1. For frequency bin $k$, the power as a function of second Fourier transform frequency, $f^\prime$, is written as
\begin{equation}
Z_k(f^\prime) = \frac{2}{\langle \lambda(f^\prime)\rangle}\left|\mathcal{F}\left[\widetilde{P}_k^n\right]\right|^2\,,
\label{eq:2ndFFT}
\end{equation}
where $\langle \lambda(f^\prime)\rangle$ is the average expected background of noise of the second Fourier transform, and $\mathcal{F}$ denotes a Fourier transform. The distribution of $Z$ from a random, Gaussian time-series is a $\chi^2$ distribution with two degrees of freedom due to the mean subtraction and smallness of correlations introduced by 50\% overlapping SFTs in equation~\ref{eq:Pmaster}. Additionally, the Central Limit Theorem can be applied because the Fourier transform in equation~\ref{eq:2ndFFT} sums random variables with finite variance, resulting in a distribution that approaches a $\chi^2$ distribution with two degrees of freedom in the large sum limit.

When there is a persistent Doppler modulated signal, the values of $Z$ show excess power in $f^\prime$ frequency bins corresponding to harmonics of the binary orbital period. The goal is then to efficiently find those values of $Z$ with excess power and characterize any signals that may be present, or to set upper limits if no signals are found.

The noise background, $\lambda(f^\prime)$, of $Z(f^\prime)$ must be characterized in order to determine the false alarm probability of candidate signals. To determine $\lambda(f^\prime)$, the expectation value of each SFT's background powers is computed across the band of interest, $E[\langle P_k\rangle]^n$. Then, drawing from an exponential distribution for each SFT with $E[\langle P_k\rangle]$ defining the distribution, a simulation of equation~\ref{eq:Pmaster} is created assuming purely Gaussian noise (including the effect of correlations due to 50\% overlapping, Hann-windowed SFTs). Next, $Z(f^\prime)$ is calculated from this simulation. This process is repeated many times (typically 200 trials) and the values for $Z(f^\prime)$ are averaged to find an estimate of the background noise power, $\lambda(f^\prime)$, for each sky location. Then, $\lambda(f^\prime)$ is scaled for different SFT frequencies across the band of interest, depending on the root-mean-square value of the time series of powers in those SFT frequency bins with respect to the average across the frequency band.

\subsection{TwoSpect detection statistic}\label{sec:Rstat}
Assume the signal power is distributed among $M$ pixels of the second Fourier transform for a narrow band of SFT frequencies, with the fraction of the signal in pixel $m$ equal to $w_m$. A useful statistic to sum pixel powers is
\begin{equation}
 R = \frac{\sum_{m=0}^{M-1} w_m(x_m - \lambda_m)}{\sum_{m=0}^{M-1}w_m^2}\,,
\label{eq:Rstat}
\end{equation}
where $x_m$ is the second Fourier transform power in pixel $m$, and $\lambda_m$ is the expected noise value of pixel $m$ of the second Fourier transform (see section~\ref{sec:dataprep}). Each value of $m$ is unique for frequency bins $k$ and $f^\prime$. For noise-only data, the expectation value of $R$ will equal zero by design. If the input time series of data is random, Gaussian white noise, then the value of $R$ is a weighted $\chi^2$ variable with up to $2M$ degrees of freedom with zero-mean. The weights, $w_m$, for each $m$ are determined by using a set of templates with parameters $(f,P,\Delta f)$ using the same $T_{\rm SFT}$ and $T_{\rm obs}$ as the search (see section~\ref{sec:templates}).

Since the power spectrum of a time series of Fourier powers is used by the $R$ detection statistic, the value of $R$ is proportional to the amplitude of the strain signal to the fourth power. We expect the value of the reconstructed strain amplitude, $h_{\rm rec}$, to scale with the value of $R$, $T_{\rm SFT}$, and $T_{\rm obs}$ by
\begin{equation}
 h_{\rm rec} \propto \left(\frac{R}{T_{\rm SFT}T_{\rm obs}}\right)^{1/4}\,.
 \label{eq:hproptoR}
\end{equation}
Thus, for increasing observation time and given a threshold for which signals are detectable at a particular confidence  level, the detectable strain amplitude decreases as the fourth-root of the observation time and the SFT coherence time.

The scale factor to convert equation~\ref{eq:hproptoR} into an equality is determined using a series of simulated signals with random frequencies, periods, and modulation depths. The mean value of these scale factors is used to determine the relationship between $R$ and $h_{\rm rec}$ for a circularly polarized signal and leads to
\begin{equation}
 h_{\rm rec} \simeq 3\, S_h^{1/2}\left(\frac{R}{T_{\rm SFT}T_{\rm obs}}\right)^{1/4}\,,
 \label{eq:hrec}
\end{equation}
where $S_h^{1/2}$ is the noise amplitude spectral density.

\subsection{Computation of templates}\label{sec:templates}
Template weights are calculated using two methods. The first method is called ``Gaussian'' because the second Fourier transforms are calculated from a series of periodic Gaussian pulses in the time domain for which analytic calculation proves tractable. These pulses give a simplified, convenient representation of the periodic power versus time pattern seen in SFT powers for a fixed frequency bin $k$. This train of Gaussian pulses is described by the equation
\begin{equation}
 x_k(t) = \sum_{n=1}^N \left[e^{-(t-nP)^2/2\sigma^2}+e^{-(t-nP-\Delta_k)^2/2\sigma^2}\right]\,,
 \label{eq:gaussianTS}
\end{equation}
where $k$ signifies the first Fourier transform frequency bin, $N = {\rm round}(T_{\rm obs}/P)$, $\Delta_k$ is the characteristic time between adjacent Gaussian pulses in a single orbit in frequency bin $k$, and $\sigma$ defines the width of the Gaussian functions. Equation~\ref{eq:gaussianTS} mimics the frequency change of the periodically varying signal in a binary system.

For convenience, we compute the continuous Fourier transformation of $x_k(t)$
\begin{equation}
 x_k(\omega) = \sqrt{2\pi\sigma^2} \sum_{n=1}^N \left[e^{-\omega(2{\rm i}nP+\omega\sigma^2)/2} + e^{-\omega[2{\rm i}(nP + \Delta_k)+\omega\sigma^2]/2} \right]\,.
\end{equation}
Taking the magnitude squared, we recover the powers:
\begin{equation}
 |x_k(\omega)|^2 = 4\pi\sigma^2 e^{-\sigma^2\omega^2} [1 + \cos(\Delta_k\omega)] \frac{\cos(NP\omega)-1}{\cos(P\omega)-1}\,,
 \label{eq:gaussianPowers}
\end{equation}
where $\omega$ is the Fourier transform variable. To use equation~\ref{eq:gaussianPowers} for a signal with expected values of $(f,P,\Delta f)$, the values of $\Delta_k$ and $\sigma$ must be computed. The value of $\Delta_k$ is readily determined to be
\begin{equation}
 \Delta_k = \frac{P}{2}-\frac{P}{\pi}\sin^{-1}\left[\frac{f_k-f_0}{\Delta f}\right]\,,
\end{equation}
where in the case that $|f_k-f_0|/\Delta f>1$, we set $\Delta_k = 0$. The rate at which the signal moves from one frequency bin to the next in the first set of Fourier transforms determines the size of $\sigma$ (the higher the rate, the smaller the value of $\sigma$). This relationship is numerically determined from simulations of various signal velocities.

The second method of template weight calculation is called ``exact'' because the templates used are closer to the true numerical values of a potential signal in each pixel of the second Fourier transform than the Gaussian templates. Note, however, that these are not the {\it true} values of the signal power for a given set of signal parameters, merely better approximations than the Gaussian template calculation method. The advantage of the Gaussian templates is that the weights are analytically determined; so the calculation of the weights is computationally efficient. In the case of the exact templates, the power value for each SFT frequency bin is analytically determined, assuming a signal with parameters $(f,P,\Delta f)$ (and an arbitrary initial phase); then the Fourier transform of each SFT frequency bin is computed. The additional steps are more computationally costly and so are used only at the final stages of the pipeline in an all-sky search.

The power of a signal-only, Hann-windowed, normalized Fourier transform is given by
\begin{equation}
 |x_z(n)|^2 = \frac{2}{3} \frac{\textrm{sinc}^2[k + \Delta fT_{SFT}\sin(2\pi nT_{1/2}/P) - z]}{\left[(k + \Delta fT_{SFT}\sin(2\pi nT_{1/2}/P) - z)^2-1\right]^2}\,,
\end{equation}
where $\textrm{sinc}(x)\equiv\sin(\pi x)/(\pi x)$, $n$ is the SFT number, $k$ is the SFT frequency bin for a non-modulated signal, $z$ is a particular SFT frequency bin for which the power is to be computed, and $T_{1/2}$ is the midpoint of the SFT coherence time. For practical purposes, when $|k + \Delta fT_{\rm SFT}\sin(2\pi nT_{1/2}/P) - z|>5$ the power is sufficiently small to set $|x_z|^2$ to zero. Once each SFT is analytically computed, the FFT for each frequency bin $z$ is computed to create the template.

For either template calculation method, the largest $M$ weight values are normalized by the sum of all weights computed. The distribution of weights is dependent on the signal parameters. To achieve maximum sensitivity, every weight for a given template should be used, but this increases the computations required. Using a smaller number of weights is more efficient. The number of pixels that sum to 90\% of the total template weight is typically of the order of tens to hundreds of pixels. The largest weights and their pixel locations are identified and sorted in descending order.

A clustering algorithm is used in the TwoSpect pipeline in order to (1) recombine multiple candidates related to a single source into a single, most significant candidate, and (2) reduce computational costs in later pipeline stages. Clustering is accomplished by grouping candidates into sequences of signal frequency and nearby binary orbital periods. Then, the range of frequency modulation amplitude values are tested and the most significant candidate is selected.

The final results of the TwoSpect pipeline are reconstructions of the signal parameters $(h_{\rm rec},f_{\rm rec},P_{\rm rec},\Delta f_{\rm rec})$ for candidates passing the threshold levels for each sky position. Hence for a measured signal using only the TwoSpect analysis, we can gain insight into only the amplitude of the gravitational wave signal $h$, binary orbital parameter $P$, and gravitational wave frequency $f$. Since we are measuring $\Delta f_{\rm obs}$ and not $\Delta f_{\rm max}$, we are unable to separate the mass ratio $q$ from the binary orbital inclination angle $i$.

\subsection{Placement of templates}\label{sec:templateplacement}
\begin{figure}
    \begin{center}
	\includegraphics[width=0.9\textwidth]{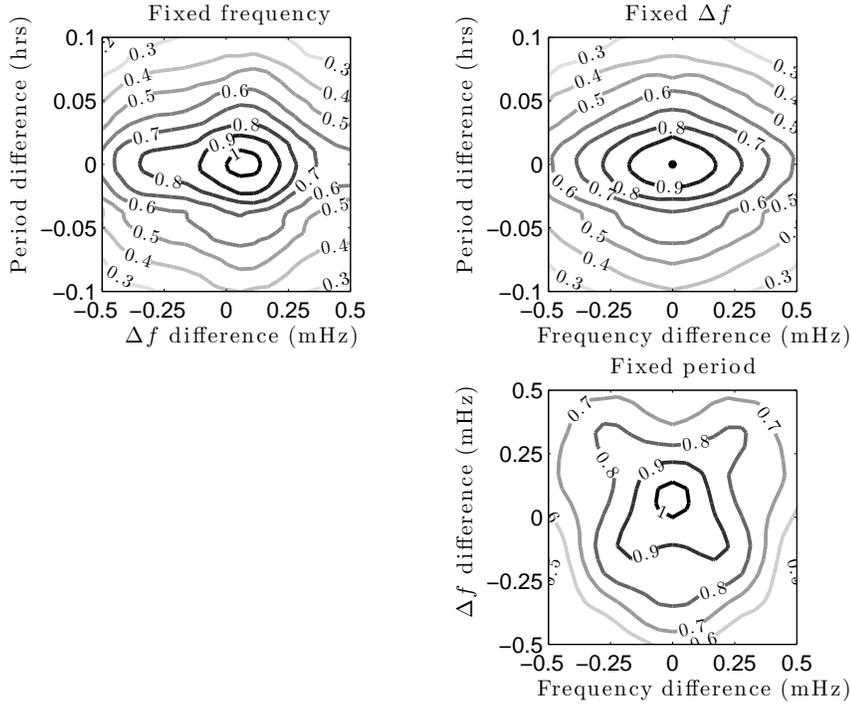}
	\caption{Contour curves of $R/R_0$ from different ``exact'' templates matched against an injected signal with parameters $f=100$ Hz, $P=14.274$ hrs, and $\Delta f=3.67$ mHz. In each plot, one parameter is held fixed while varying the remaining parameters.}
	\label{fig:countourcurve}
    \end{center}
\end{figure}

\begin{figure}
    \begin{center}
	\includegraphics[width=0.7\textwidth]{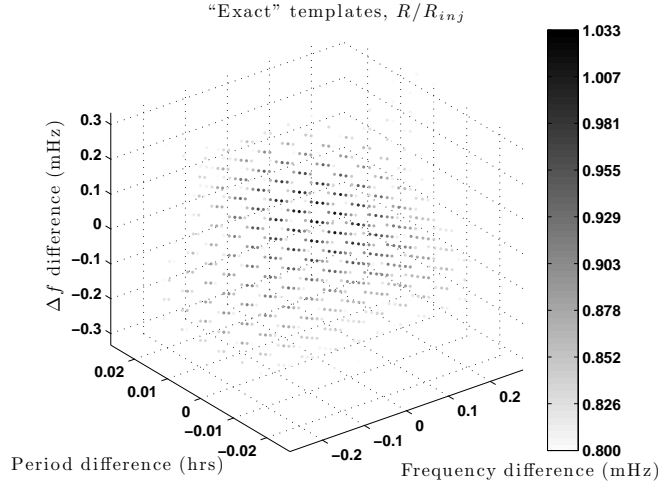}
	\caption{Scatter plot of $R/R_0$ from different ``exact'' templates matched against an injected signal with parameters $f=100$ Hz, $P=14.274$ hrs, and $\Delta f=3.67$ mHz. Here, only templates with a normalized $R$ value greater than or equal to 0.8 were kept. This corresponds to a mismatch 0.2 or less.}
	\label{fig:3dscatteroftemplates}
    \end{center}
\end{figure}

Computational limitations restrict the amount of parameter space that can be covered using a lattice of templates placed for a given allowed mismatch $\mu$. The mismatch defines the fraction by which $R$ is reduced when using a template that does not match the true signal parameters. The maximum separation values were determined empirically using simulated data. Figures~\ref{fig:countourcurve} and \ref{fig:3dscatteroftemplates} show the results of one set of simulations to compute template spacings. Figure~\ref{fig:countourcurve} shows a wide range of template trials, holding a single search parameter fixed while varying the remaining two parameters. In figure~\ref{fig:3dscatteroftemplates}, a three-dimensional plot shows the range of those templates which yield a relative value of $R/R_0\geq0.8$ where $R_0$ is the value of $R$ from a template given the true parameters.

Placing templates in frequency space would give a mismatch of $\lesssim$0.2 for templates spaced by $1/(2T_{\rm SFT})$ Hz. This indicates that the resolution of the signal frequency using a mismatch value of 0.2 would be no better than $1/(2T_{\rm SFT})$ Hz. Decreasing the mismatch value, i.e. decreasing spacing between templates, would improve the resolution but increase the computational cost of the search.

For placement of the templates in binary orbital period we use an iterative routine:
\begin{equation}
 \Delta P = P_1 - P_0
\end{equation}
where $P_1$ is the new period a distance $\Delta P$ away from the previous period $P_0$. This can be written assuming that the signal is shifted some fraction of a second Fourier transform frequency bin,
\begin{equation}
 \Delta P = \frac{1}{1/P_0 - 1/(\alpha T_{\rm obs})} - P_0\,.
\end{equation}
Here, $1/(\alpha T_{\rm obs})$ is the fractional bin shift for a given mismatch $\mu$, and $\alpha$ is an empirically derived parameter. To first order in $P_0/(\alpha T_{\rm obs})$,
\begin{equation}
 \Delta P \simeq \frac{P_0^2}{\alpha T_{\rm obs}}\,.
\end{equation}
The empirically derived value for $\alpha$ is linearly dependent on the coherence time as $\alpha\simeq2.7(T_{\rm SFT}/1800\,{\rm s})+1.8$. The empirically derived value for $\Delta P$ scales inversely with the square root of the modulation depth because the signal at the turning points of the frequency variation in the second Fourier transform scales with the square root of the modulation depth for a fixed binary orbital period. Therefore, the distance in period spacing between templates must be reduced as the modulation depth increases.

The spacing in modulation depth is similar to the spacing of templates in frequency. For a mismatch of $\mu\lesssim0.2$ the spacing of templates is about $1/(2T_{\rm SFT})$.

Using the above expressions, it is possible to determine the expected number of templates needed for a particular set of search parameters. For example, given $\mu=0.2$, the number of templates for the binary orbital period is
\begin{eqnarray}
\fl N_P(\Delta f) \simeq \left[1.17\times10^4\left(\frac{T_{\rm SFT}}{1800\,{\rm s}}\right) + 7.71\times10^3\right] \left(\frac{T_{\rm obs}}{1\,{\rm yr}}\right) \left(\frac{P_{\rm min}}{2\,{\rm h}}\right)^{-1} \left(\frac{\Delta f}{3.6\,{\rm mHz}}\right)^{-1/2}\,.
\end{eqnarray}
Here, it is assumed that $P_{\rm min}\ll T_{\rm obs}$ and $P_{\rm min}\ll P_{\rm max}$. Note that the number of templates has a power-law dependence on the modulation depth of the signal, where calculated values were determined with $\Delta f=3.6$ mHz. The number of templates per sky location in the 3-dimensional parameter space $(f_0,P,\Delta f)$ is
\begin{equation}
 N_{f,P,\Delta f} = \int_{f_{\rm min}}^{f_{\rm max}}\int_{\Delta f_{\rm min}}^{\Delta f_{\rm max}} \frac{N_P(\Delta f)}{{\rm d}\Delta f {\rm d}f} \,{\rm d}\Delta f {\rm d}f\,.
\end{equation}
Taking ${\rm d}\Delta f = {\rm d}f = 1/(2T_{\rm SFT})$, the double integral can be evaluated. Searching the whole sky in a 1 Hz band at 100 Hz, covering a range of periods down to 2~h, and using 1 year of data broken into 1800~s segments would require $N_{\rm tot}=N_{\rm sky}N_{f,P,\Delta f}$ templates,
\begin{eqnarray}
\fl N_{\rm tot} \simeq \left[6\times10^{15}\left(\frac{T_{\rm SFT}}{1800\,{\rm s}}\right) + 4\times10^{15}\right] \left(\frac{T_{\rm obs}}{1\,{\rm yr}}\right) \left(\frac{T_{\rm SFT}}{1800\,{\rm s}}\right)^4 \left(\frac{P_{\rm min}}{2\,{\rm h}}\right)^{-1} \left(\frac{f_{\rm band}}{1\,{\rm Hz}}\right) \nonumber \\
 \times\left(\frac{f}{100\,{\rm Hz}}\right)^2 \left[\left(\frac{\Delta f_{\rm max}}{3.6\,{\rm mHz}}\right)^{1/2} - \left(\frac{\Delta f_{\rm min}}{3.6\,{\rm mHz}}\right)^{1/2}\right]\,.
\end{eqnarray}
Since this number is so large and computationally demanding, we choose, in the all-sky search, to apply a hierarchical approach to avoid explicitly searching over every template. This hierarchy is defined by a pre-template stage that reduces the parameter space to be searched with the templates (see figure~\ref{fig:twospectpipeline}). In contrast, using TwoSpect to carry out a search for a source at a known sky location is computationally tractable without the hierarchical structure.

\subsection{Incoherent harmonic sum}\label{sec:ihs}
The all-sky search begins with an untemplated search algorithm, incoherent harmonic summing (IHS)~\cite{LongFFTTechniques}, to identify regions of parameter space to be searched later using templates. In this algorithm, each $Z_k(f^\prime)$ is stretched an integer $j=1\ldots S$ times, each stretched spectra is summed, and the maximum value is chosen according to
\begin{equation}
\mathcal{V}_k = \textrm{max}\left[\sum_{j=1}^SZ_k(f^\prime/j)\right]\,.
\end{equation}
If a periodic signal is present, then the IHS algorithm will accumulate the signal into the harmonic frequencies of the signal. The signal-to-noise ratio of the signal bins grow $\propto$$\sqrt{S}$, provided the harmonic powers have similar SNR in the original spectra. In practice, this increase in SNR is limited by the strength of the signal harmonics, giving the IHS technique a practical limit of $S\sim5$ in this application. 

\begin{figure}
 \begin{center}
  \subfigure[Noise-only.]{\includegraphics[width=0.49\textwidth]{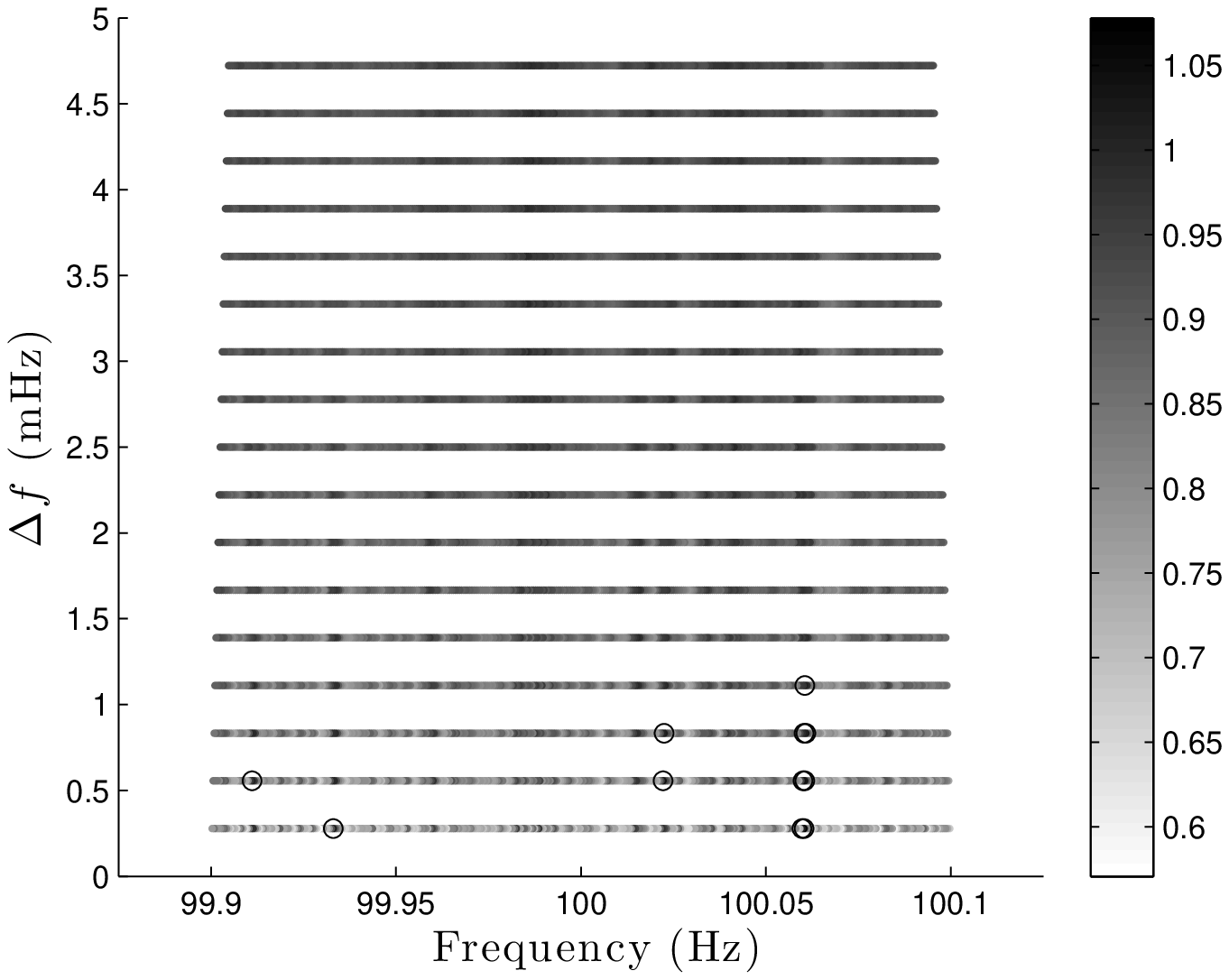}}
  \subfigure[Moderate strength simulated signal injection.]{\includegraphics[width=0.49\textwidth]{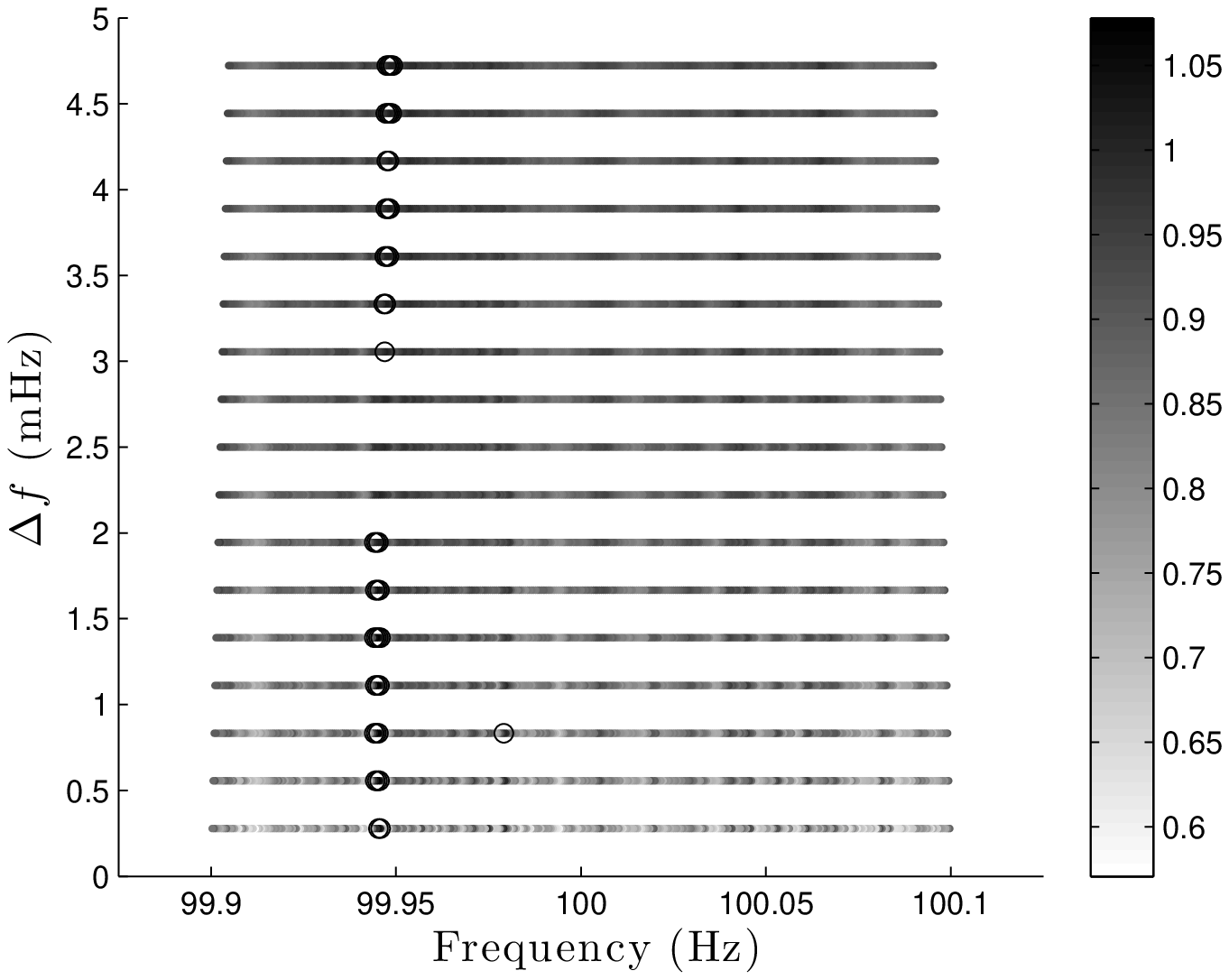}}
  \caption{The maximum IHS values, $\mathcal{V}_k$, across sequential SFT frequency bins, $k$ normalized by the expected false alarm threshold value for frequency modulation amplitudes between 0.5 and 9.5 frequency bins ($0.28\leq \Delta f\leq 5$~mHz). Open circles mark candidates above the threshold. (a) The noise-only case shows a few candidate events. Note the variance of the $\mathcal{V}_k$ values decreases with increasing modulation depth due to the sum across multiple SFT bins. (b) The moderate strength signal case shows a few candidate events (indicated by black empty circles). Note the correlations along the vertical axis.}
  \label{fig:IHSplots}
 \end{center}
\end{figure}

To accumulate additional signal power, $Z_k$ values are summed across sequential values of $k$ before computing $\mathcal{V}_k$. Each series of $\mathcal{V}_k$ values is subjected to a threshold test (see figure~\ref{fig:IHSplots}). Then, a coincidence test must be satisfied, where the most significant pair of single $\mathcal{V}_k$ values in the sequence of SFT frequencies must be symmetric across the series of single $\mathcal{V}_k$ values.

The coincidence criterion is determined using a $\chi^2$ test for symmetry of a signal in the $Z(f^\prime)$ domain. Let $P$ be the number of row pairs, $m_i^l$ and $m_i^u$ be the measured locations in $f^\prime$ of $\mathcal{V}_k$ for the lower and upper value in $k$ in pair $i$, $\sigma_{l_i}$ and $\sigma_{u_i}$ be the uncertainties in measured locations, and $T_i$ is the true location of the $\mathcal{V}_k$ values (this enforces the symmetry). The statistic $\mathcal{M}$ is a $\chi^2$ test for symmetry
\begin{equation}
 \mathcal{M} = \sum_{i=1}^P\left[\frac{(m_i^l-T_i)^2}{\sigma_{l_i}^2} + \frac{(m_i^u-T_i)^2}{\sigma_{u_i}^2}\right]\,.
\end{equation}
Subject to minimization, it reduces to
\begin{equation}
 \mathcal{M} = \sum_{i=1}^P \frac{1}{\sigma_{l_i}^2 + \sigma_{u_i}^2}(m_i^l-m_i^u)^2\,.
\end{equation}

Since the location of $\mathcal{V}_k$ is simply a single bin in row $k$, the variance in this value is $1/12$. The variance is then weighted by the signal-to-noise ratio, $\mathcal{S}$, of $\mathcal{V}_k$. Substituting $\sigma^2 = 1/(12\mathcal{S})$ we find,
\begin{equation}
 \mathcal{M} = \sum_{i=1}^P \frac{12 \mathcal{S}_{l_i} \mathcal{S}_{u_i}}{\mathcal{S}_{l_i} + \mathcal{S}_{u_i}}(m_i^l-m_i^u)^2\,.
\end{equation}
In practice, the largest contribution to $\mathcal{M}$ will be dominated by the pair that has the highest combined SNR, we therefore determine $\mathcal{M}$ for only the highest combined SNR pair.

The threshold $\mathcal{V}_k$ value is determined by a Monte Carlo simulation of exponentially distributed second Fourier transform noise with expectation values determined by $\lambda(f^\prime)$. The IHS algorithm is applied to the simulated noise and threshold levels computed for the number of sequential $Z_k$ values summed. Next, the detector data is compared with the threshold levels determined from Monte Carlo simulations. Once candidate regions of parameter space have passed threshold tests using the IHS routine, the candidate signals are subjected to a threshold test on $R$ using templates based on the values found from the IHS step as described in section~\ref{sec:templates}.

\subsection{Significance of candidate events}\label{sec:candsig}
Candidates are characterized by their false alarm probability, that is, the probability of the candidate's $R$ value arising in a signal-free sample of Gaussian noise. In random noise alone, the TwoSpect search statistic $R$ is the weighted sum of $\chi^2$ random variables, each with 2 degrees of freedom but with differing expectation values for each variable. For equally weighted variables, the distribution of the sum approaches a Gaussian distribution in the limit of a sum of infinite variables. In the other extreme, when only one variable dominates, the distribution of the weighted sum is approximately exponential.

The problem of calculating the false alarm probability for a sum of weighted $\chi^2$ variables is well known in statistics. The probability that a value of $R$ exceeds a threshold value of $R_0$ for a sum of $N$ weighted, $\chi^2$ random variables with two degrees of freedom, with each random variable having expectation value $\lambda_i$ and an associated weight, $w_i$ is given by
\begin{equation}
 P(R\geq R_0) = \sum_{i=1}^N\frac{e^{\frac{-R_0}{w_i\lambda_i}}}{\prod_{j\neq i}\left(1-\frac{w_j\lambda_j}{w_i\lambda_i}\right)}\,,
 \label{eq:probSumNweigtedChiSqVariables2}
\end{equation}
and is similar in form to equation~8.4 in~\cite{AllenCWstrategies}. Unfortunately, equation~\ref{eq:probSumNweigtedChiSqVariables2} diverges as any number of weights times expectation values approach similar values.

It proves useful to use a {\it characteristic function} to determine the probability of exceeding a threshold in a different way. This technique converts the probability distribution function of random variables to the Fourier domain. Characteristic functions can be used to find the probability distribution for a weighted sum of random $\chi^2$ variables which is discussed in detail in~\cite{ImhofNormVars,Davies1}.

\begin{figure}
 \begin{center}
  \subfigure[]{\includegraphics[width=0.433\textwidth]{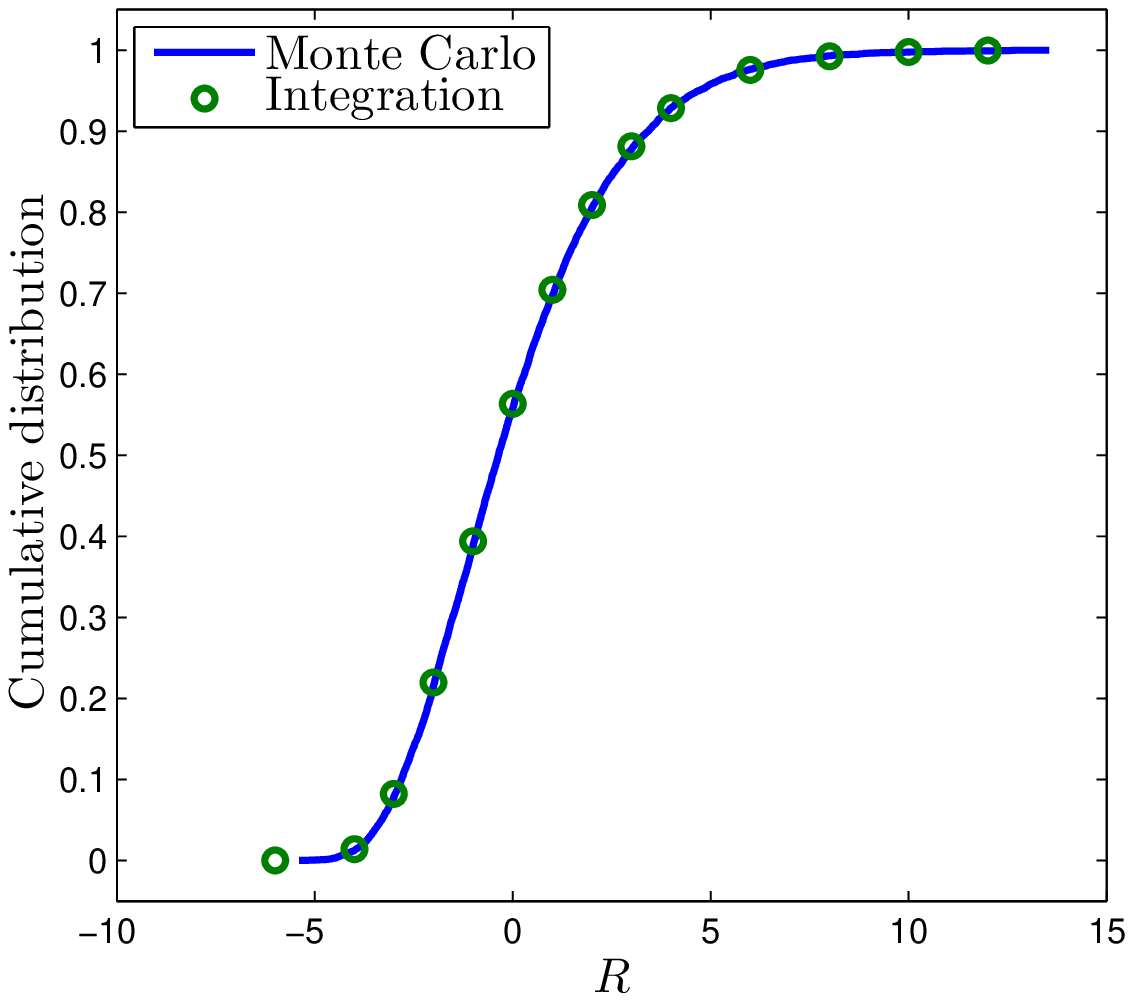}}
  \subfigure[]{\includegraphics[width=0.45\textwidth]{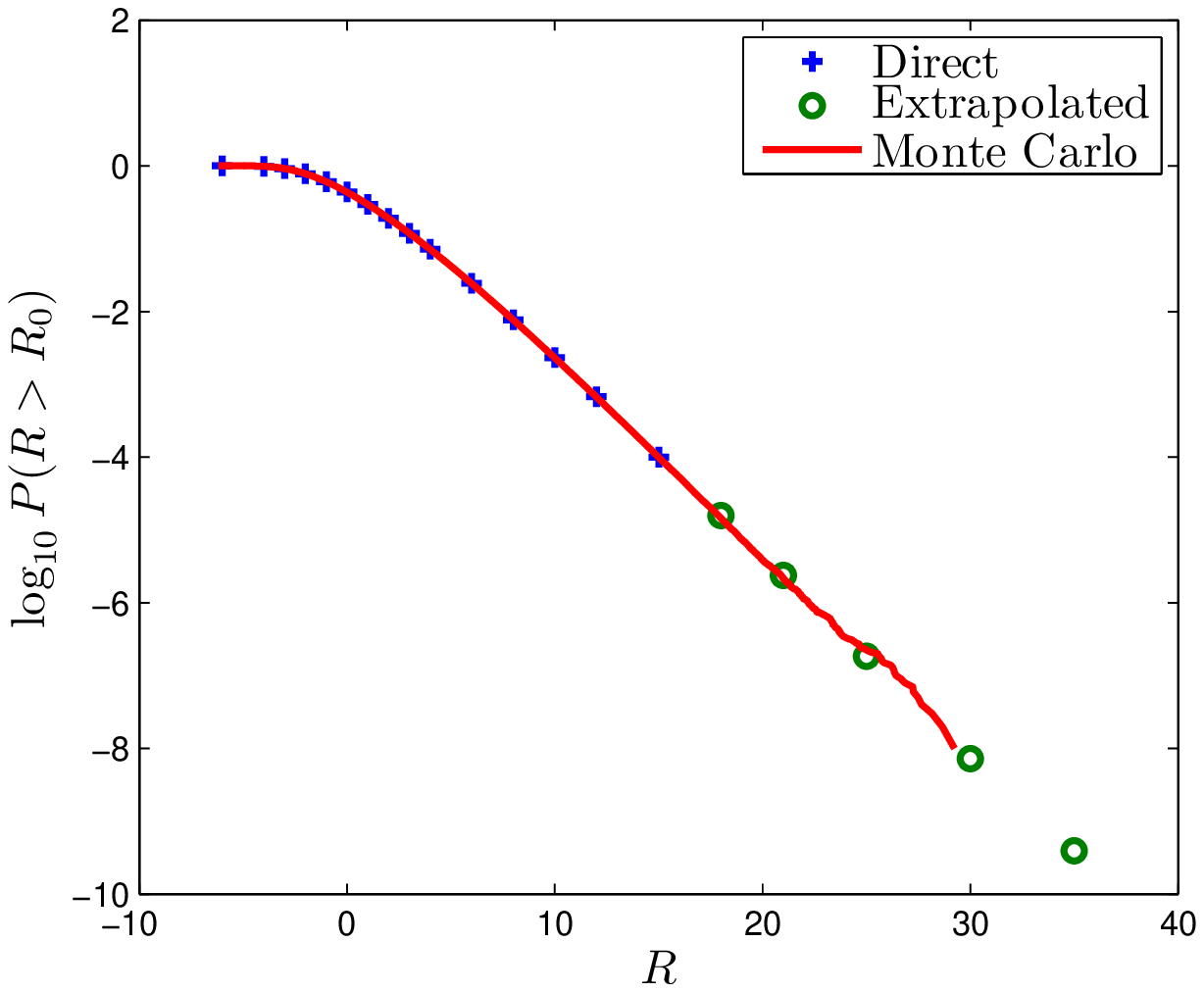}}
  \caption{(a) The cumulative distribution function of $R$ for a particular template of $w_i$ values determined using Monte Carlo simulations over Gaussian noise (line) and by numerical integration of the Gil-Pelaez formula (circles). (b) Extrapolation of rare events (circles) by exploiting the linear logarithmic probability function in the regime of rare events (tail of crosses). A Monte Carlo simulation using exponentially distributed random variables (line) confirms the validity of the extrapolation.}
  \label{fig:cdfrealexpected}
 \end{center}
\end{figure}

Since each of these random variables in the sum, equation~\ref{eq:Rstat}, is independent, the characteristic function of $R$ (neglecting mean subtraction, which is simply a rescaling factor) will take the form
\begin{equation}
 \phi_R(u) = \prod_{j=1}^N\frac{1}{1-{\rm i}u w_j^\prime\lambda_j}\,.
\end{equation}
In the case of TwoSpect, the weights and expectation values are independent for each random variable. Determining the probability that $R$ lies {\it below} a value $R_0$ is then given by the Gil-Pelaez formula~\cite{Gil1951}
\begin{equation}
 P(R<R_0) = \frac{1}{2} - \int_{-\infty}^{\infty} \Im\left(\frac{\phi_R(u)e^{-{\rm i}uR}}{2\pi u}\right) \,du
 \label{eq:gilpelaez}
\end{equation}
and is related to the probability of exceeding the threshold by $P(R\geq R_0) = 1-P(R<R_0)$. Solving equation~\ref{eq:gilpelaez} requires numerical integration. The details of the integration method can be found in~\cite{Davies2}. Although this integral is not solved analytically, the performance of this numeric calculation is significantly faster than estimating the probability function using Monte Carlo simulations. Figure~\ref{fig:cdfrealexpected}(a) compares the results of Monte Carlo simulation with the numerical integration routine, and figure~\ref{fig:cdfrealexpected}(b) shows the technique (direct integration) being used to extrapolate to the rare event regime.

The solution to this integral can be used in two ways. First, it can be used when a candidate signal has been detected for a given set of weights, expected noise values for given pixels, and a value of $R_0$ to determine the probability that purely random noise values would have produced a value of $R$ that is as large as the found $R_0$. Second, this equation can be solved iteratively for $R_0$ when we wish to set a particular false alarm probability threshold value.

\subsection{Running the TwoSpect analysis code}
The TwoSpect program is written in C and stored in the LSC (LIGO Scientific Collaboration) Algorithms Library (LALapps) repository~\cite{LALrepository}. TwoSpect reads in previously stored SFTs calculated from the calibrated detector $h(t)$ channel for sliding, weighting, and computation of the second Fourier transform (see figure~\ref{fig:twospectpipeline}). The code performs all necessary calculations and outputs a list of candidates that have exceeded threshold values.

The parameter space defined in section~\ref{sec:paramspace} is divided into narrow spans of frequency and submitted as separate, parallel jobs which are run on LSC computer clusters under the Condor environment. Each job searches over the the entire sky for a range of frequency, binary orbital period and signal modulation depth parameters.

The outer loop of the code searches over sky position. Then the inner loop of the incoherent harmonic sum step-searches over signal frequency and modulation depth, identifying possible binary orbital periods associated with initial candidates. Next, each candidate identified in the incoherent harmonic sum step is passed to the template-based algorithm. Each candidate is compared with possible templates. Those that pass threshold tests are considered to be outliers meriting follow-up studies (see section~\ref{sec:twospectresults}).

\section{Validation of the TwoSpect pipeline}\label{sec:twospectresults}
In addition to testing the various functions of the TwoSpect program to verify algorithm correctness, an end-to-end validation of the complete TwoSpect pipeline is carried out by two types of simulations. First, simulated detector data consisting of {\it pure noise} has been used to determine threshold levels to limit the number of outliers and estimate the typical noise background (see section~\ref{sec:purenoisetwospect}). Second, as described in section~\ref{sec:injectionstwospect}, the pipeline has been subjected to different software injections using simulated data with various signal parameters and strain amplitudes in order to determine its sensitivity. These validations have confirmed the promising potential and robustness of the pipeline.

\subsection{Simulated noise-only data}\label{sec:purenoisetwospect}
A ten-week sample of simulated detector noise-only data has been generated using the Makefakedata program. This LALApps repository~\cite{LALrepository} program was run with options set to create Hann-windowed, noise-only 1800~s SFTs with 50\% overlap between each SFT. The SFT data was centered at 100~Hz, with a bandwidth of 0.2~Hz, and had frequency bin spacing $\delta f = T_{\rm SFT}^{-1} \simeq 5.556\times10^{-4}$~Hz. The noise was random, Gaussian, and stationary, with an expectation value of the amplitude spectral density set (for testing convenience) equal to 0.023~57~Hz$^{-1/2}$\footnote{The typical noise amplitude spectral density of an initial or enhanced 4-km LIGO detector near 100~Hz is of the order of $10^{-22}\,{\rm Hz}^{-1/2}$.}. The range of parameter values for this search was $f_0=[99.9,100.1]$~Hz, $P=[2,336]$~h, and $\Delta f=[0.27,98.3]$~mHz. The false alarm rate for the IHS step in this run was set at 0.1\% while the threshold rate for the template steps was set at $0.01$\% in order to further reject false signals. In the later stages of the pipeline, the templates were placed with a mismatch of $\mu\approx0.2$ as discussed in section~\ref{sec:templateplacement}. A template-based search only (e.g., a search not using the IHS algorithm) would require of order $10^{15}$ templates to cover this parameter space with the defined mismatch.

\begin{figure}
 \begin{center}
  \includegraphics[width=0.8\textwidth]{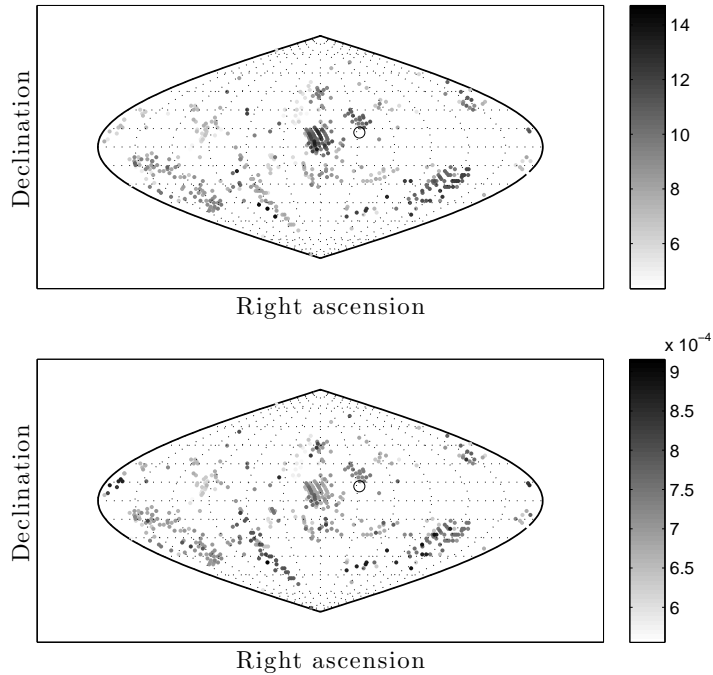}
  \caption{Sky maps of candidates surviving threshold tests using the TwoSpect algorithm from a time series of Gaussian noise. The upper plot are the candidates' negative logarithmic false alarm probabilities. The lower plot are the candidates' reconstructed strain amplitude assuming circular polarization. The black circles in each plot indicates the average position of the Sun during the observation time. Zero hours right ascension is located at the right of the plot, with increasing right ascension as one moves to the left.}
  \label{fig:simdataskymapProbAndH0}
 \end{center}
\end{figure}

The search described above required about 13~h on a single computing node of an LSC computer cluster. Figure~\ref{fig:simdataskymapProbAndH0} shows sky maps of the candidates surviving the threshold cuts used. The upper map shows the resulting logarithmic false alarm probabilities for each candidate's $R$ value, and the lower map shows each candidate's reconstructed strain amplitude, assuming circular polarization, using equation~\ref{eq:hrec}. The candidates are expected to be randomly scattered over the sky with sporadic arcs around regions where the noise, by random chance, has higher values and the Doppler shift between grid points is not large. These arcs are concentric with the average position of the Sun during short time observations and define regions with comparable magnitudes of the projection of the Earth's average acceleration toward the Sun during that time span, giving comparable expected Doppler modulation patterns~\cite{S4PSH}. The corresponding candidates are dominated by templates with small frequency modulation depths. The number of IHS ``templates'' searched is of the order of $10^9$. The IHS templates are not strictly independent since there exist correlations between neighboring SFT frequency bins when the IHS values from the different bins are summed (see figure~\ref{fig:IHSplots}). The false alarm threshold calculations include these correlations when computing the threshold limits.

\begin{figure}
 \begin{center}
  \subfigure[]{\includegraphics[width=0.49\textwidth]{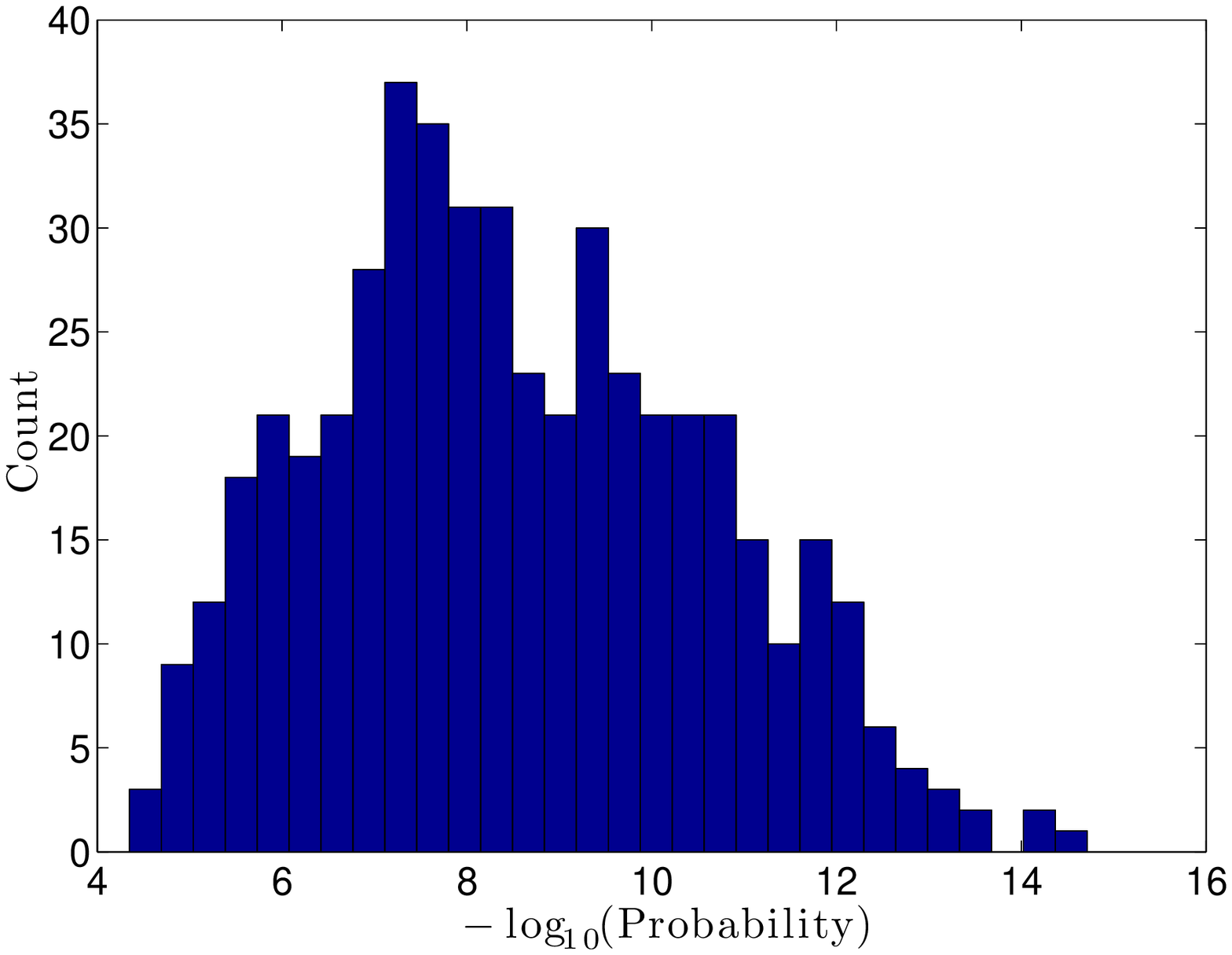}}
  \subfigure[]{\includegraphics[width=0.49\textwidth]{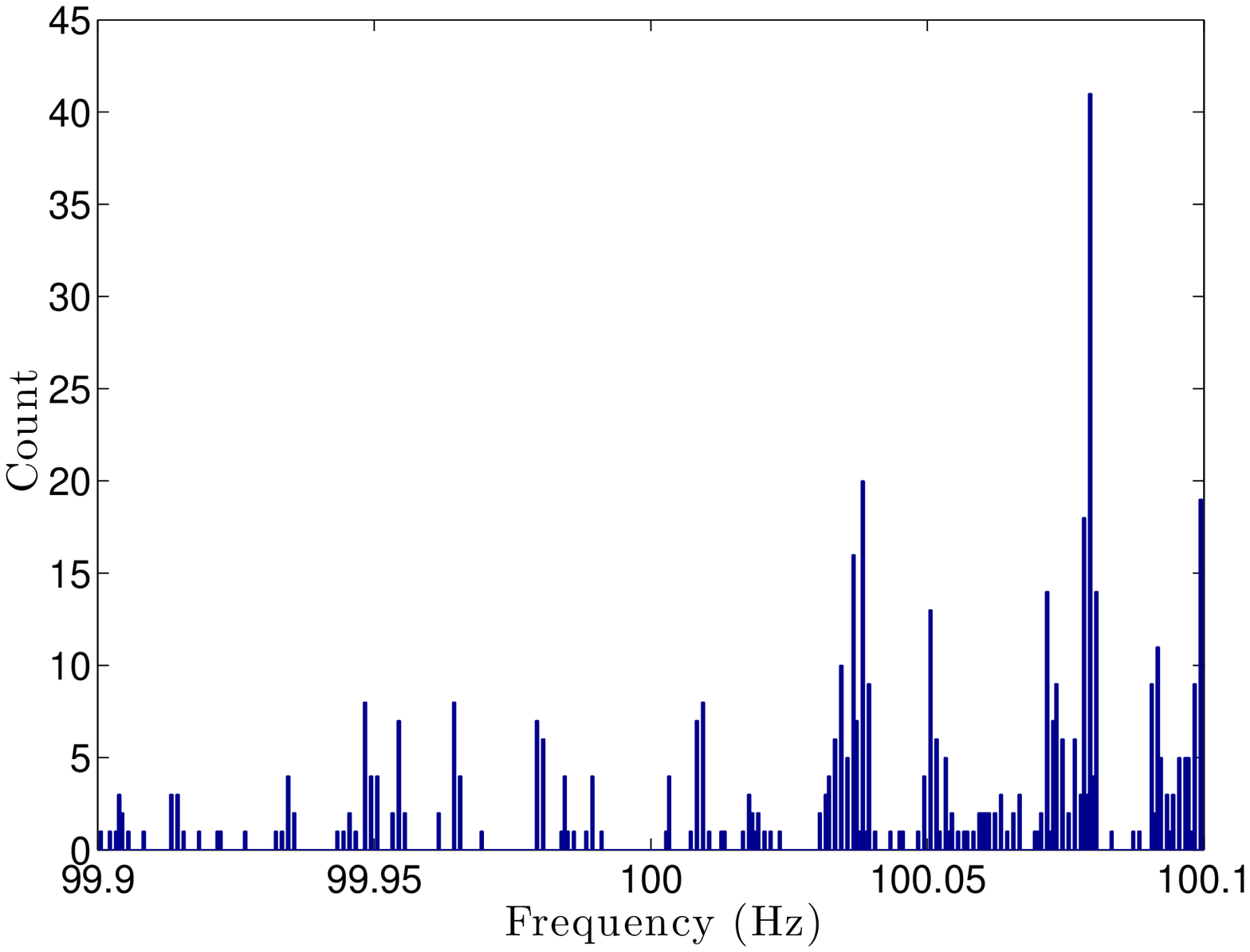}}\\
  \subfigure[]{\includegraphics[width=0.49\textwidth]{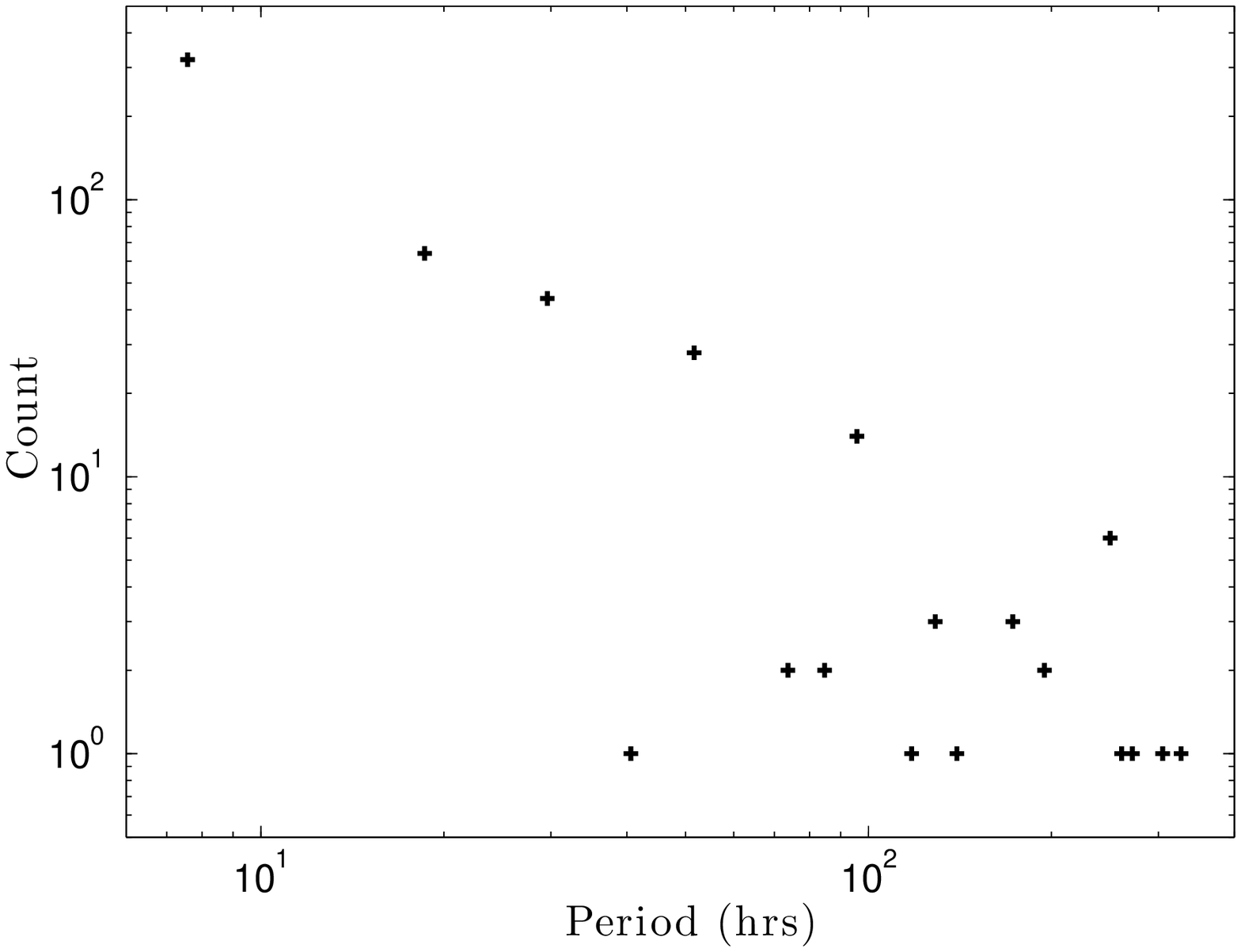}}
  \subfigure[]{\includegraphics[width=0.49\textwidth]{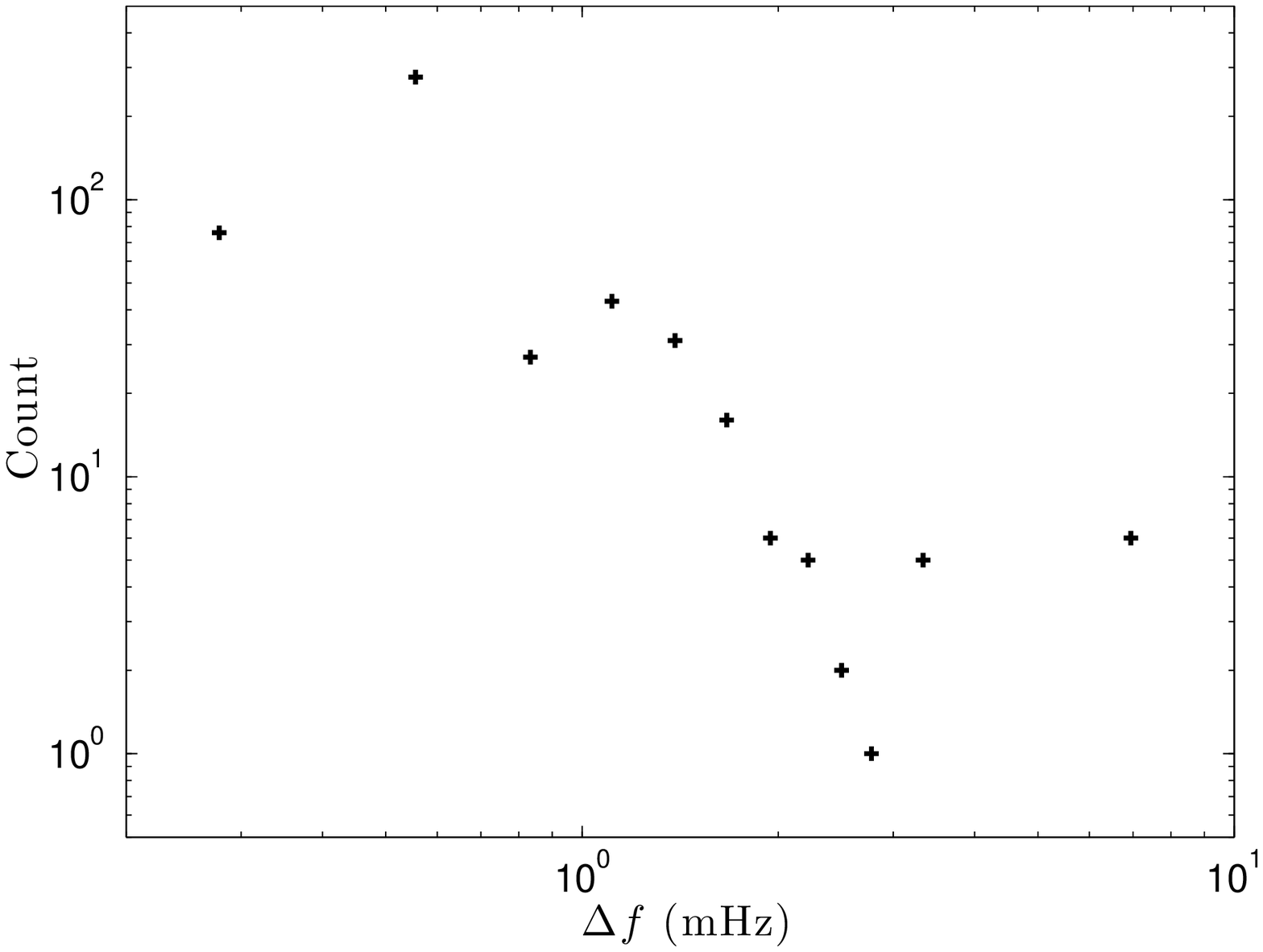}}
  \caption{Distributions of the candidates found using noise-only data with the IHS false alarm rate set at 0.1\% and the template false alarm rate set at $10^{-2}$\%. Histograms are of the noise-only candidates' (a) $-\log_{10}$ false alarm probability, (b) gravitational wave frequency, (c) period of the hypothetical binary orbit, and (d) amplitude of frequency modulation.}
  \label{fig:simdatahists}
 \end{center}
\end{figure}

Histograms of the noise-only candidates' parameters are shown in figures~\ref{fig:simdatahists}(a)--(d). The candidates are distributed non-uniformly in frequency (see figure~\ref{fig:simdatahists}(b)), with clusters near SFT frequency bins that have above-average noise values in their second Fourier transforms. The distribution of candidate signals in binary orbital period parameter (see figure~\ref{fig:simdatahists}(c)) shows that short periods (high second Fourier transform frequency) occur more often in pure noise due to the increased number of templates with short periods. The increase is proportional to $P^{-2}$. Figure~\ref{fig:simdatahists}(d) shows that all candidates from this simulation have low modulation depths ($<$10~mHz) which correspond to modulations of only a small number of SFT frequency bins ($<$18 bins). Excess noise in a few SFT frequency bins is more likely to cause such spurious candidates.

It is observed that some of the outliers seen in the noise-only data have extremely small false alarm probabilities (see figure~\ref{fig:simdatahists}(a)). Uncertainty in the background estimate can have an effect on the calculated false alarm probability that is reported for each candidate if the background is systematically underestimated for the powers in the template. Variations in the logarithmic probability values for the largest outliers have been observed at the level of $\pm2$ for variations in the parameters used in background estimation. These variations do not, however, fully explain the largest outlier values seen in the simulated noise-only data.

Residual correlations in the simulated data not included in the expected background estimates are a likely cause of these loud apparent outliers, which appear to stem from modulation of the underlying noise envelope by antenna pattern weighting. Accounting for these correlations to
produce more accurate probability estimates is under investigation.

A future improved version of the TwoSpect pipeline will implement a coincidence test between the different interferometers in order to remove spurious outliers and keep threshold levels for the pipeline stages low. The criteria for coincidence will require consistency among the source parameters of $(\alpha,\delta,f,P,\Delta f)$ found for corresponding candidates in each interferometer. Imposing these constraints should greatly reduce the number of loud outliers for detailed follow-up and allow reduction of single-interferometer threshold levels. Simulation studies to guide coincidence criteria are under way.

\subsection{Signal injection recovery}\label{sec:injectionstwospect}
The TwoSpect pipeline has also been tested with fake signals of various source parameters, including different strain amplitudes, created using the Makefakedata program. Table~\ref{tab:simsig} shows the set of simulated signals and their source parameters. Test data for pulsar numbers 1 through 10 was generated for the Hanford 4-km interferometer (H1) starting at GPS time 900000000, lasting for 10 weeks of total observation time, with zero spin-down, and cosine of the star's inclination angle $\cos\iota=1.0$ (yielding circularly polarized waves). The projected semi-major axis was set in each case to provide the Doppler shift indicated, and the orbital eccentricity was set to zero. The noise amplitude spectral density, $\sqrt{S_h}$, was set equal to 1.0~Hz$^{-1/2}$ in every case, with the same noise used in every test by setting the same random noise seed value for every simulation.

Table~\ref{tab:simsig} also shows the corresponding recovered pulsar signal parameters for injections 1 through 10. The TwoSpect algorithm is given the correct sky location of the injection, but no other parameters are given. The search is performed over the entire band with the only restriction on $\Delta f_{\rm rec}$ given by equation~\ref{eq:TsftLim}. Injections 1 through 5 are loud enough that the pipeline correctly identifies them. Injections 6 through 10 are detected with reduced significance ($\log_{10}$(Prob.) nearer to zero) and the recovered signal parameters are less accurate than the strong signals, as discussed below. Only the most significant candidate for each searched sky location is listed in table~\ref{tab:simsig}.

The low-amplitude recovered signals, specifically injection numbers 6 through 10, have inaccurate reconstructions of the true signal. The typical identified parameters that are incorrect are the signal frequency and the frequency modulation depth, especially as the signal is spread over more SFT frequency bins with increasing modulation depth. Collectively, injections 6 through 10 have smaller $h_0$ values and greater $\Delta f$ values leading to degraded accuracy in signal reconstruction. At low SNR, TwoSpect has substantially better precision on the period, $P$, than on the source frequency, $f$, or modulation depth, $\Delta f$.

Next, four simulated signals from spinning neutron stars in elliptical orbits were generated. Test data for pulsar numbers 11 through 14 (see table~\ref{tab:simsig}) was again created for the H1 detector starting at GPS time 900000000, lasting for 10 weeks of total observation time, with zero spin-down, cosine of the inclination angle $\cos\iota=1.0$, time of periapsis passage at 900000000.0 (the time of closest approach between the two stars in the SSB frame), argument of periapsis equal to 0.0 radians (this parameter defines the rotation of the elliptical orbit on the sky), and the orbital eccentricity parameter ranging from $10^{-4}$ to $10^{-1}$.

Table~\ref{tab:simsig} shows the corresponding recovered pulsar signal parameters from neutron stars 11 through 14 in elliptical orbits. The TwoSpect algorithm is given the correct sky location of the injection, but no other parameters are given. The search is again performed over the entire 0.2~Hz band, with the only restriction, on $\Delta f_{\rm rec}$, given by equation~\ref{eq:TsftLim}. Only the most significant candidate for the sky location is listed in table~\ref{tab:simsig}. Currently, the pipeline does not search over orbital eccentricity of the binary system, so no value of orbital eccentricity is reconstructed.

As shown in table~\ref{tab:simsig}, the eccentric orbit signals are recovered by the TwoSpect algorithm with nearly the correct frequency, binary orbital period, and reconstructed strain values. The reconstructed modulation depth is approximately correct for small eccentricity, but for larger eccentricity values, the modulation depth measures only the amplitude of the Doppler shifted frequency. Thus, for modest orbital eccentricity, the TwoSpect algorithm is still able to recover signals.

\begin{figure}
 \begin{center}
  \includegraphics[width=1.0\textwidth]{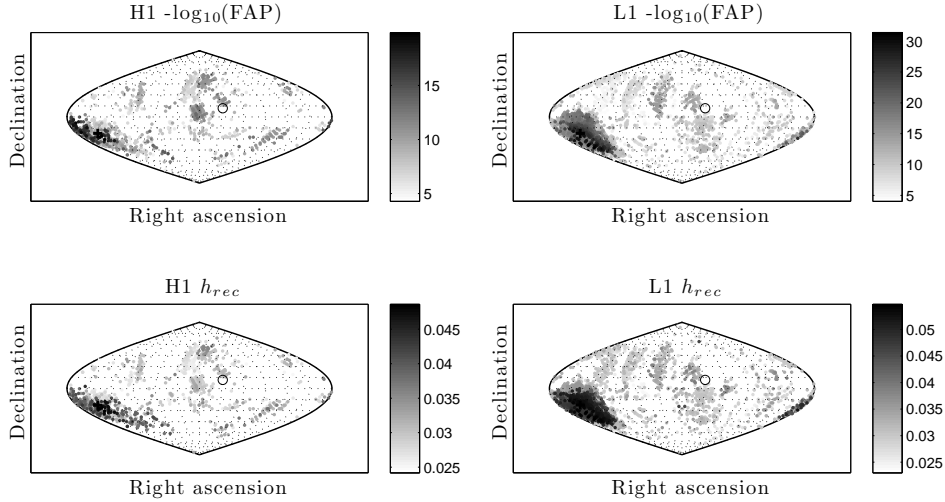}
  \caption{Sky maps of a simulated signal in the H1 and L1 LIGO 4~km interferometers. Candidates surviving threshold tests using the TwoSpect algorithm are shown with their negative logarithmic false alarm probability (upper plots) and reconstructed strain amplitude assuming circular polarization (lower plots).  Zero hours right ascension is located at the right of the plot, with increasing right ascension as one moves to the left. The average position of the sun during the observation time is shown as a black circle while the position of the source is given by a black cross.}
  \label{fig:inj9skymapProbAndH0}
 \end{center}
\end{figure}

To illustrate how a potential signal would be seen as a sky map in two different gravitational wave detectors (the LIGO 4~km interferometers H1 and L1), two separate data streams are generated with the same signal but with different noise in each stream, as well as different antenna pattern functions (for each interferometer) applied to each signal. Injection number nine's sky location and orbital parameters are used, but with $h_0=0.04$. Each data stream has noise spectral density $\sqrt{S_h} = 1.0$, and is analyzed separately using the TwoSpect pipeline. Figure~\ref{fig:inj9skymapProbAndH0} shows the results for the logarithmic false alarm probability (upper plots) and the reconstructed strain amplitude (lower plots) for each candidate passing threshold tests.

Figure~\ref{fig:inj9skymapProbAndH0} demonstrates how two widely separated gravitational wave interferometers would observe a potential signal. For this simulation, the L1 interferometer is more sensitive to the signal than the H1 interferometer due to its more favorable antenna pattern coverage (by 15\% in strain sensitivity) for this sky region. The result is an improved ability to distinguish a signal from the surrounding noise.

\section{Conclusions and outlook}
Direct detection of continuous gravitational waves from a spinning neutron star would not only add to the understanding of General Relativity, but also provide valuable constraints on neutron star equations of state. A number of searches for particular sources and all-sky searches for unknown isolated neutron stars have been undertaken over the past decade, but no previous search has been conducted for unknown neutron stars in binary systems. The additional search parameters of the binary system make the methods adapted from the isolated all-sky routines computationally intractable.

The TwoSpect algorithm will enable searches for unknown neutron stars in binary systems in gravitational wave detector data. Since more than half of the known pulsars in the LIGO frequency band (assuming gravitational wave emission occurs at twice the rotational frequency of the neutron star) are in binary systems, this method has potentially great application to the neutron star population of our galaxy. A detection of a continuous wave signal would be groundbreaking in the field of gravitational wave physics. Although the search does not attain the strain sensitivity of other all-sky search methods for isolated neutron stars (e.g., PowerFlux, Einstein@Home), it is the first all-sky algorithm now searching the region of parameter space for quasi-monochromatic gravitational waves emanating from previously unknown neutron stars in binary systems. (Another all-sky binary search method, named Polynomial, is under active development and will use a polynomial matched-filter approach with coherence times shorter than orbital periods~\cite{PolynomialSearch}.)

Compared to the searches for isolated stars with known ephemerides, all-sky searches generally have a degradation by at least an order of magnitude in their upper limits due to the computational limitations imposed by such searches~\cite{S4PSH}. Additionally, the current all-sky search algorithms are not designed to cope with the increase in the computational costs search the binary orbital parameter space. For example, the PowerFlux algorithm~\cite{S4PSH} would require of order $10^{21}$ templates or more to search the same 0.2~Hz frequency band as described in section~\ref{sec:purenoisetwospect}. A purely template-based TwoSpect search would require of order $10^{15}$ templates. A search method such as PowerFlux is much more sensitive than the TwoSpect algorithm once the binary orbital parameters are known. As an example, in the case of Injection 1 (see table~\ref{tab:simsig}), the PowerFlux search statistic would find a candidate signal with a Gaussian signal-to-noise ratio (SNR) of about 51, about double the SNR of the most significant TwoSpect candidate at the same sky location. For this reason, the potential gain in SNR is useful for follow-up studies of candidate signals using other search algorithms.

The essential methods used in the TwoSpect algorithm have been validated, and the initial version of the pipeline has been implemented in C code which is now running on LSC computer clusters. End-to-end tests have been conducted using simulated noise-only data and using individual simulated signals from neutron stars in binary systems in noisy data. These tests demonstrate the pipeline's readiness to begin a full search in LIGO science run 6 (S6) and Virgo's second and third science runs (VSR2 and VSR3) data in 2011.

\ack
We wish to thank our colleagues in the LIGO-Virgo Continuous Waves working group for helpful discussions, especially B. Allen, V. Dergachev, C. Messenger, R. Prix, and K. Wette. This work has been supported by the National Science Foundation under grant number NSF-PHY0855422. This article has LIGO document number LIGO-P1100021.

\section*{References}
\bibliography{refs}

\begin{landscape}
\Table{\label{tab:simsig}Summary of simulated data signals used to test the TwoSpect pipeline, and the loudest recovered signals at the true sky location of the injected source assuming circular polarization. The final column, $-\log_{10}$(FAP), is the negative logarithm of the false alarm probability.}
  \begin{tabular}{@{}llllllllllllll}
   \br
    & $\alpha$ & $\delta$ & $a\sin i$ & Ecc. & & & $f$ & $f_{\rm rec}$ & $P$ & $P_{\rm rec}$ & $\Delta f$ & $\Delta f_{\rm rec}$ & $-\log_{10}$ \\
   \# & (hrs) & (deg.) & (s) & $e$ & $h_0$ & $h_{\rm rec}$ & (Hz) & (Hz) & (hrs) & (hrs) & (mHz) & (mHz) & (FAP) \\
    \mr
    1 & \00.0 & \00.0 & \0\00.3  & 0 & 0.06 & 0.06 & 100.0000 & 100.0001 & \014.274 & \014.270 & \03.668 & \03.611 & 121.9 \\
    2 & \03.050 & \-15.59 & \0\00.0232 & 0 & 0.05 & 0.05 & 100.0091 & 100.0092 & \0\05.013 & \0\05.008 & \00.808& \00.833 & \084.0 \\
    3 & 21.921 & 23.83 & \0\09.3530 & 0 & 0.07 & 0.08 & 100.0815 & 100.0815 & 152.173 & 151.642 & 10.736 & 10.833 & 153.9 \\
    4 & 19.207 & \-64.46 & \013.041 & 0 & 0.08 & 0.10 & 100.0028 & 100.0028 & \091.876 & \092.035 & 24.775 & 24.722 & 280.5 \\
    5 & 15.121 & 43.30 & \0\01.9596 & 0 & 0.08 & 0.10 & \099.9504 & \099.9504 & \054.400 & \054.501 & \06.284 & \06.389 & 682.1 \\
    6 & 11.691 & \0\-9.29 & 137.7776 & 0 & 0.07 & 0.04 & 100.0330 & 100.1092 & 312.317 & 311.710 & 77.020 & \00.556 & \021.2 \\
    7 & \08.750 & 78.88 & \041.3526 & 0 & 0.05 & 0.06 & \099.9670 & \099.9611 & 261.442 & 262.315 & 27.597 & 33.611 & \021.7 \\
    8 & 22.556 & 18.718 & \0\00.1529 & 0 & 0.04 & 0.04 & \099.9146 & \099.9147 & \010.950 & \010.945 & \02.435 & \02.500 & \031.7 \\
    9 & 21.962 & \-22.412 & \0\02.4814 & 0 & 0.06 & 0.05 & 100.0426 & 100.0478 & \033.509 & \033.448 & 12.930 & \07.222 & \043.2 \\
    10 & \02.046 & \-50.373 & \027.5270 & 0 & 0.05 & 0.06 & 100.0150 & 100.0028 & 117.309 & 117.887 & 40.961 & 28.333 & \021.8 \\
    11 & 21.921 & 23.83 & \0\09.353 & $10^{-4}$ & 0.07 & 0.08 & 100.0815 & 100.08155 & 152.173 & 151.845 & \0\0--- & 10.556 & 146.3 \\
    12 & 21.921 & 23.83 & \0\09.353 & $10^{-3}$ & 0.07 & 0.08 & 100.0815 & 100.08151 & 152.173 & 150.938 & \0\0--- & 10.833 & 157.3 \\
    13 & 21.921 & 23.83 & \0\09.353 & $10^{-2}$ & 0.07 & 0.08 & 100.0815 & 100.08143 & 152.173 & 151.491 & \0\0--- & 10.833 & 136.4 \\
    14 & 21.921 & 23.83 & \0\09.353 & $10^{-1}$ & 0.07 & 0.08 & 100.0815 & 100.08044 & 152.173 & 151.807 & \0\0--- & 10.833 & 149.6 \\
    \br
\end{tabular}
\endTable
\end{landscape}

\end{document}